\documentclass{aa}  
\usepackage{titlesec}
\titleformat{\subparagraph}[runin]{\normalfont\normalsize\bfseries}{\thesubparagraph}{1em}{}
\usepackage{microtype}
\usepackage[varg]{txfonts}
\usepackage{graphicx}
\usepackage{mathrsfs}
\usepackage{placeins}
\usepackage{makecell}
\usepackage{soul}
\usepackage{overpic}
\usepackage{tikz}
\usepackage{float}
\usepackage{amsmath}
\usepackage{lscape}  
\usepackage{subcaption}

\usepackage{txfonts}
\usepackage{svg}
\usepackage[colorlinks=true, linkcolor=blue, citecolor=blue, urlcolor=blue]{hyperref}
\usepackage{natbib}
\usepackage{titlesec}

\usepackage[normalem]{ulem}

\begin{document} 
\sloppy

   \title{The O\,Vz stars in NGC~346: distribution, stellar parameters and insights into low-metallicity effects}

   \author{L. Arango\inst{1}
          \and
          J. I. Arias \inst{1}
          \and
          G. Holgado\inst{2} 
          \and
          G. Ferrero\inst{3,4}
          \and
          C. Putkuri\inst{3}\
          \and
          N. I. Morrell\inst{5}
          }

   \institute{Universidad de La Serena, Departamento de Física y Astronomía,
              Av. Raúl Bitrán 1305, La Serena, Chile\\
              \email{johanna3193@gmail.com}     
         \and
             Instituto de Astrofísica de Canarias, E-38200 La Laguna, Tenerife, Spain  
        \and
             Instituto de Astrofísica de La Plata, CONICET-UNLP, Paseo del Bosque s/n, B1900FWA, La Plata, Argentina
        \and
             Facultad de Ciencias Astronómicas y Geofísicas, Universidad Nacional de La Plata, Paseo del Bosque s/n, B1900FWA, La Plata, Argentina     
         \and
           Las Campanas Observatory. Carnegie Observatories. Casilla 601. La Serena, Chile     
           }


  \abstract 
   {O\,Vz stars are identified by optical spectra where the $\rm{He\,II\,\lambda4686}$ absorption line is more intense than any other He line, suggesting they may be less luminous, less evolved, and closer to the zero-age main sequence (ZAMS). Despite ongoing debate, these stars remain candidates for understanding the early evolutionary stages of O-type stars.}
   {We study O\,Vz stars in the star-forming regions NGC 346 in the Small Magellanic Cloud (SMC) and 30~Doradus in the Large Magellanic Cloud (LMC), by determining their physical parameters and exploring the influence of low metallicity on the development of the Vz spectral peculiarity.}
   {We identified a sample of O\,Vz and O\,V stars in NGC 346 using spectra obtained with the Magellan Echellette (MagE) spectrograph at Las Campanas Observatory. The stars were classified based on the relationship between the equivalent width of the relevant He lines. Quantitative spectroscopic analysis was performed using the IACOB-BROAD, IACOB-GBAT/FASTWIND tools. The O\,V and O\,Vz stars from 30~Doradus, previously identified in an earlier study, were reclassified following the same criteria.}
   {In NGC~346, 8 O\,Vz stars (29\%) and 20 O\,V stars (71\%) were identified. In 30~Doradus after a revised classification, 28 O\,Vz stars (33\%) and 56 O V stars (67\%) were identified. Despite differences in sample sizes and metallicity, the proportion of O\,Vz stars is similar in both regions. O\,Vz stars in NGC 346 exhibit lower projected rotational velocities ($v \, \mathrm{sin} \, i$ < 70 $\rm{km\,s^{-1}}$), higher effective temperatures (37–40 kK) and similar surface gravities ($\log {\mathrm{g}}$ = 3.7–4.1) compared to O\,V stars. Contrary to prior assumptions, O\,Vz stars are not consistently closer to the ZAMS. In the Hertzsprung-Russell (HR) diagram, O\,Vz stars appear in both young and evolved regions, suggesting that the Vz phenomenon is related to stellar wind properties.}
   {}

   \maketitle

\section{Introduction}

Massive stars play a fundamental role in galactic evolution, yet our understanding of their formation and early evolution remains limited. These stars are challenging to study because they are often located in distant regions with significant extinction and exhibit extraordinary properties, such as intense stellar winds \citep{kudritzki2000winds, bestenlehner2014vlt, hawcroft2024x}, significant mass loss \citep{lucy2012star, puls2008, vink2017mass}, magnetic fields \citep{walder2012magnetic,wade2017magnetism} and high luminosities. They are also relatively rare compared to lower-mass stars \citep{salpeter1955luminosity} and have much shorter \mbox{lifespans}, typically ranging from $5$ to $20$ Myr \citep{ekstrom2012grids, langer2012presupernova}. Studying these stars at the youngest possible stages is \mbox{essential}, as they may retain evidence of their formation processes.

Over the years, studies of O-type dwarfs in the optical domain, both in the Milky Way (MW) and the Magellanic Clouds (MCs), have provided valuable insights into the physical properties and evolutionary stages of massive stars. Among these, a luminosity subclass known as O\,Vz stars has been identified. O\,Vz stars were first identified in the Galactic cluster Trumpler 14 \citep{walborn1973some} and later in the Magellanic Clouds \citep{parker1992stellar, walborn1992two, walborn1997spectral}. This subclass was introduced by \cite{walborn1997spectral} and is defined by an enhanced absorption in the $\rm{He \,II \,\lambda}$$4686$ line compared to other helium lines. This feature has been hypothesized to indicate lower luminosities and higher surface gravities, suggesting that O\,Vz stars are closer to the zero-age main sequence (ZAMS). While the Vz peculiarity has the potential to identify extremely young massive stars, only a limited number of such objects have been analyzed through quantitative spectroscopy \citep{sabin2014vlt, holgado2020iacob, rickard2022stellar}.

Subsequent studies revealed that the interpretation of the Vz phenomenon is more complex, requiring the consideration of multiple factors beyond stellar winds \citep{sabin2014vlt, arias2016spectral}. Detailed spectroscopic analyses and \mbox{atmospheric} models have shown that the appearance of the Vz feature in a stellar spectrum may depend on a specific combination of stellar parameters, including effective temperature, surface gravity, projected rotational velocity, and multiplicity. Moreover, in a systematic study of Galactic O-type stars from the Galactic O Star Spectroscopic Survey \citealp[GOSSS]{sota2011galactic, sota2013galactic, apellaniz2016galactic}, \citet{arias2016spectral} quantitatively \mbox{characterized} the Vz phenomenon by analyzing the equivalent widths of the $\rm{He \,I \,\lambda}$$4471$, $\rm{He \,II \,\lambda}$$4542$, and $\rm{He \,II \,\lambda4686}$ spectral lines. This work refined the criterion initially proposed by \citet{sota2014galactic}, making it more stringent by assigning the ``z'' \mbox{qualifier} only to spectra where the equivalent width of $\rm{He \,II \,\lambda4686}$ is at least 1.10 times the maximum equivalent width of either $\rm{He \,I \,\lambda}$$4471$ or $\rm{He \,II \,\lambda}$$4542$, depending on which of the two is stronger at the given spectral type.

The Magellanic Clouds, with metallicities of approximately 1/2 (LMC) and 1/5 (SMC) of the solar value, provide an ideal laboratory to investigate the effects of metallicity on massive star evolution and phenomena like the Vz classification. NGC 346 in the SMC is the most O-star-rich star-forming region in the galaxy, housing nearly half of its known O-type stars \citep{massey1989stellar}. With an age of less than 3 Myr \citep{bouret2003quantitative}, NGC 346 is one of the youngest and most massive clusters in the Local Group, offering a unique opportunity to study massive stars in a low-metallicity environment. The most comprehensive spectroscopic studies of this cluster to date include those by \citet{dufton2019census}, who identified a very young population \mbox{(<2 Myr)}, and \citet{rickard2022stellar}, who applied the quantitative Vz classification criterion of \citet{arias2016spectral} to identify one potential Vz star in the region.

30 Doradus (30 Dor), in the LMC, is the most massive star-forming region in the Local Group \citep{walborn2014vlt}. It hosts the most massive stars known \citep{crowther2010hot,bestenlehner2011vlt} and serves as a reference for massive star studies at low metallicities. \cite{sabin2014vlt} analyzed O Vz stars in 30 Dor and found that, although these stars are generally closer to the ZAMS, many have ages between 2 and 4 Myr. This result suggests a potential connection between metallicity and the persistence of the Vz phase. 

In this article, we perform a quantitative spectroscopic \mbox{analysis} of O Vz stars in NGC 346, using the Vz classification criterion proposed by \citet{arias2016spectral}. Additionally, we compare the properties of O Vz and O V stars in NGC 346 with those previously analyzed in 30 Doradus by \cite{sabin2014vlt}. Our study builds on the former findings, \mbox{providing} a low-metallicity counterpart to further investigate how the Vz phenomenon may vary between different environments. The structure of this paper is as follows: in Section.~\ref{ch:Methodology}, we describe the observations, the Vz classification criteria, and the methods used to estimate stellar parameters. The results are presented in Section.~\ref{ch:Results}, followed by a detailed discussion in Section.~\ref{ch:Discussion}. Finally, a summary of our main findings is provided in Section.~\ref{ch:Summary}.

\section{Methodology}\label{ch:Methodology} 

\subsection{Sample of study}
This study is primarily based on intermediate-resolution optical spectra obtained with the MagE spectrograph \mbox{\citep{marshall2008mage}}, mounted on the Clay 6.5m telescope at Las Campanas Observatory. The MagE data were kindly provided by Phil Massey, Kathryn Neugent, and Nidia Morrell. These spectra were complemented with public VLT/FLAMES spectra, previously analyzed by \citet{dufton2019census}. Confirmed spectroscopic binaries were excluded from the analysis.

The total sample of stars analyzed in NGC~346 comprises 37 objects. Throughout the paper, the designation \mbox{``NGC 346-\#Star''} is adopted to ensure the proper identification of individual objects. Instrumental configurations and data characteristics are summarized in Table~\ref{Table:Table_1}.

Additionally, this study incorporates a reanalysis of a sample of stars in 30~Doradus, based on spectra from the VLT-FLAMES Tarantula Survey \citep{evans2011vlt}. The spectra for 30~Doradus were originally analyzed by \citet{sabin2014vlt} [hereafter, SS14]. While their main characteristics are summarized in Table~\ref{Table:Table_1}, the reader is referred to that paper for further details. 

\FloatBarrier  
\begin{table*}[ht!]
\caption[]{Characteristics of the spectra employed in this study.}
\label{Table:Table_1}
$$
\begin{array}{p{0.10\linewidth}ccccccc}
\hline \hline
\noalign{\smallskip}
\text{Region} &\text{Telescope} &\text{Instrument} & \text{Resolving power} 
 &\text{S/N} &  \lambda \text{ Range ($\AA$)} & \text{Stars} \\
\noalign{\smallskip}
\hline
\noalign{\smallskip}
\text{NGC~346} & \text{Clay-6.5m} &\text{MagE}  & 4100 & 100-400 & 3200-9000     &31   \\
      & \text{VLT- 8.2m}& \text{FLAMES} & 7000-8500 & >100 &  3960-5071 & 6  \\
\noalign{\smallskip}
\hline
\noalign{\smallskip}
\text{30~Doradus} & \text{VLT- 8.2m} & \text{FLAMES}  & 7000-16000 & \sim150 &  3960-6817 & 84 \\
\noalign{\smallskip}
\hline
\end{array}
$$
\end{table*}

To assess the representativeness of our sample, we compared it with the Stellar Spectral Classification Catalog compiled by \citep{skiff2009vizier}, which aggregates spectral classifications from various sources, including stars in the SMC and specifically in NGC 346. Although this catalog is not a complete census, it provides a broad reference for the distribution of known O-type dwarf stars (luminosity class V) in the region,  based on studies such \citet{niemela1986young, massey1989stellar, smith1997uv, evans2006vlt, lamb2016vizier, dufton2019census}.

\begin{figure}[ht!]
\centering
\includegraphics[width=8cm, height=7cm]{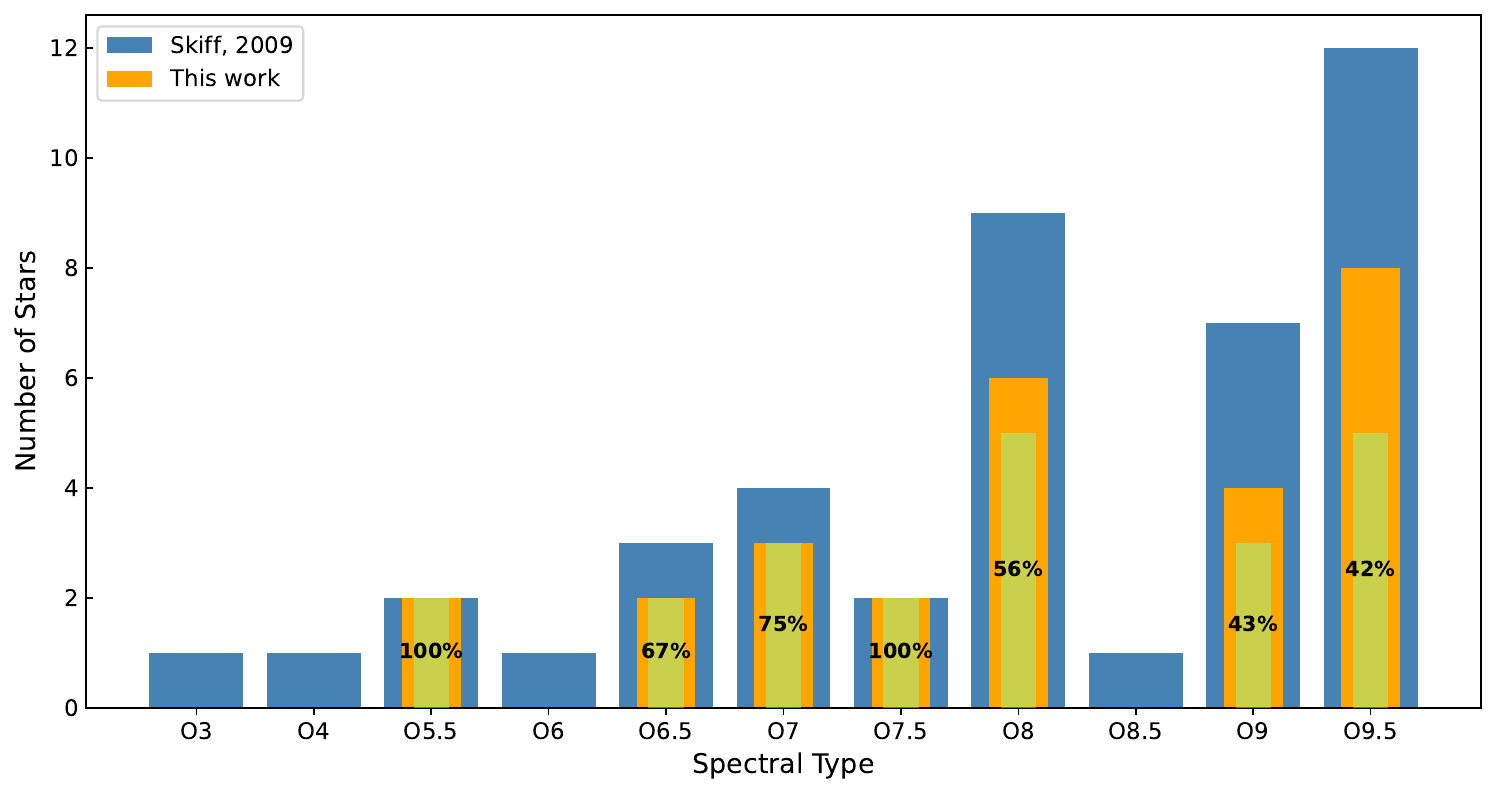}
\caption{Comparison between our sample (orange bars) of O-type stars with luminosity class V in NGC 346 and those previously reported in the literature \citep[][blue bars]{skiff2009vizier}. The green bars represent the number of stars common to both samples.  For each spectral subtype, the percentages indicate the fraction of stars in the Skiff catalogue that are also present in our sample.}
\label{Figure:Figure_1}
\end{figure}

Figure.~\ref{Figure:Figure_1} shows a histogram comparing the number of O-type dwarf stars reported in the Skiff catalog  (blue bars) with those analyzed in this work (orange bars), broken down by spectral subtype.  The green bars represent the stars common to both samples. The percentage labels in the green bars indicate, for each spectral subtype, the proportion of Skiff catalogue star present in our sample.

Although our sample was limited to the available spectra from previous observing campaigns, it spans the full range from O5.5 to O9.5 and includes a substantial fraction of stars in the most populated subtypes. This suggests that, despite its incompleteness, our sample reasonably reflects the overall spectral distribution of O-type dwarfs in NGC 346, providing a suitable basis for the analysis presented in this study.

\subsection{Spectral classification}

The spectral classifications for the stars in NGC 346, as listed in
Table~\ref{Table:Table_4}, were reviewed and updated where necessary to ensure consistency. We followed the criteria established by the Galactic O-Star Spectroscopic Survey 
\citep{sota2011galactic, arias2016spectral}. 

A key focus of this study was the classification of stars into the O\,V and O\,Vz subclasses. For this, we applied the quantitative criterion proposed by \citet{arias2016spectral} (hereafter A16), which is based on the equivalent widths (EWs) of the He I and He II lines relevant to the Vz classification.

The measurements were performed using the \texttt{splot} task in IRAF \citep[Image Reduction and Analysis Facility, ][]{tody1993iraf}, which allows interactive adjustment of spectral lines and integration of the area under the line profile. 
Errors in EW measurements were estimated based on the signal-to-noise ratio (S/N) and \mbox{instrumental} resolution. Specifically, uncertainties were calculated based on the line width at half-maximum intensity (FWHM) and continuum noise. The EW measurement and their errors are presented in Table \ref{Table:Table_3}. Since the resolving power of the instrument affects precision, the medium-resolution spectra sometimes \mbox{resulted} in overlapping or poorly defined line peaks, increasing measurement uncertainty. To validate our results, we compared the EWs obtained with IRAF to those calculated with a custom Python script, finding excellent agreement between the two methods, with minimal discrepancies.

Nebular contamination presented additional difficulties, particularly for the 
$\rm{He \,I \,\lambda}$$4471$ line, where intense nebular emission interfered with the detection of the stellar lines. Similarly, the $\rm{He \,II \,\lambda}4542$ line showed contamination in its blue wing from overlapping $\rm{N \,II \,\lambda\lambda}4511/15$ lines. In these cases, we integrated half of the line profile and doubled the value to estimate the total EW.

\subsection{Quantitative spectral analysis} \label{ch:parameters}
Our analysis strategy to estimate the stellar parameters of our sample relied on two semi-automated tools developed within the framework of the IACOB project: IACOB-BROAD \citep{diaz-herrero} and IACOB-GBAT \citep{simon2011iacob}. The latter utilizes a grid of FASTWIND models \citep{santolaya1997atmospheric, puls2005atmospheric}, 
covering a wide range of stellar and wind parameters.

We started by calculating the broadening parameters, specifically the projected rotational velocity ($v \, \mathrm{sin} \, i$) and the macroturbulence  ($v_{\mathrm{mac}}$), using IACOB-BROAD. These parameters, which significantly influence the shape of stellar line profiles, were then used as input to determine additional stellar parameters with IACOB-GBAT. The analysis focused on the optical wavelength range ($4000-5000$ $\AA$), allowing us to derive effective temperature ($T_\mathrm{eff}$), surface gravity ($\log{\mathrm{g}}$), the wind strength parameter  \citep[$\mathrm{Q}$, ][]{puls1996star}, helium abundance ($Y(\mathrm{He})$), microturbulence (\rm{$\xi_t$}).

From the diverse set of FASTWIND grids incorporated in IACOB-GBAT, we selected models calculated for $Z=0.2$ Z$_{\odot}$, corresponding to the approximate metallicity of the SMC. The grid was constructed assuming the spectroscopic parameters are treated as free adjustable variables. The parameter ranges covered by the FASTWIND model grid are summarized in Table~\ref{Table:Table_2}.

We did not perform stellar parameter determinations for the 30~Dor objects included in this study. Instead, we adopted the results from SS14, who determined the stellar parameters using the same approach employed here, based on the IACOB tools.

\begin{table}[ht!]
\caption[]{Parameter space covered by the grid FASTWIND models at SMC metallicity.}
\label{Table:Table_2}
\centering
\begin{tabular}{lcc}
    \hline \hline
    \noalign{\smallskip}
    \text{Parameter}  & \text{Range} & \text{Step size} \\
    \noalign{\smallskip}
    \hline
    \noalign{\smallskip}
    \text{\rm{Z}}               & $0.2$ Z$_{\odot}$          &  \\ 
    \text{$T_\mathrm{eff}$~\mbox{[K]}}         & $[19000, 61000]$  & $1000$  \\
    \text{$\log {\mathrm{g}}$~\mbox{[dex]}}  & $[2.1, 4.5]$    & $0.1$ \\
    \text{\rm{$\xi_t$}~\mbox{[km s$^{-1}$]}}      & $[1-30]$        & $5$ \\
    \text{$Y(\mathrm{He})$} \tablefootmark{a}            & $0.04-0.30$          &   $0.02$  \\
    \text{$\log {\mathrm{Q}}$} \tablefootmark{b}        & $[-11.0, -16.0]$ & $\sim 0.5$ \\
    \text{\rm{$\beta$}}               & $1.0$            &  \\  
    \noalign{\smallskip}
    \hline
\end{tabular}
\tablefoot{\tablefoottext{a} $Y(\mathrm{He})$ = $\rm{N(He)/N(H)}$,
\tablefoottext{b} $Q = \dot{M} (Rv_{\infty})^{-3/2}$}
\end{table}

\subsubsection*{Stellar parameter determination} 
{\em{Projected rotational velocity ($v \, \mathrm{sin} \, i$):}} 
the resolving power of our spectra ($\rm{R} \sim 4100$) and their limited wavelength coverage ($4000$–$5000$ $\mbox{\AA}$) impose constraints on the precise characterization of line broadening in the sample. At this resolution, $v \, \mathrm{sin} \, i$ values below $\sim$75~$\rm{km\, s^{-1}}$ (approximately the resolution limit, $\rm{c/R}$) are inherently uncertain. In cases where $v \, \mathrm{sin} \, i$ could not be \mbox{reliably} measured, we adopted a fixed value of 50 $\rm{km\, s^{-1}}$, following standard practice. This was applied to six O\,V and two O\,Vz stars.

Due to the general absence of metallic lines in our spectra, we relied primarily on He I lines to derive estimates of $v \, \mathrm{sin} \, i$. We used He I $\rm{\lambda}$4387 as a main diagnostic, as it is present in most stars and appropriate for this type of analysis. For a few stars where He I lines were weak or absent, we used He II $\rm{\lambda}$4541  as an alternative. It is important to note that, according to \citet{simon2007fourier}, the Fourier Transform (FT) method implemented in the IACOB-BROAD tool yields consistent $v \, \mathrm{sin} \, i$ measurements in O-type stars using both He I and metallic lines.

While the use of these lines can pose challenges for accurately studying very slow rotators, as discussed in \citet{ramirez2013vlt}, this limitation is less critical here due to the already restricted resolution of the data. \mbox{Moreover}, methodological approaches outlined in works such as \citet{gray2005} and \citet{simon2014iacob} provide a robust framework for disentangling the effects of rotational broadening and macroturbulence in massive stars, even under these constraints.\\

{\em{Diagnostic lines:}} to perform the analysis, we initially \mbox{considered} the following set of spectral lines: $\rm{H_\delta}$, $\rm{H_\gamma}$, $\rm{H_\beta}$, \text{He\,I} \mbox{$\rm{\lambda}$$4387$}, \text{He\,I} $\rm{\lambda}$$4471$, \text{He\,I} $\rm{\lambda}$$4713$, \text{He\,I} $\rm{\lambda}$$4922$, \text{He\,II} $\rm{\lambda}$$4200$, \text{He\,II} $\rm{\lambda}$$4541$, \text{He\,II} $\rm{\lambda}$$4686$, and \text{He\,I}+\text{He\,II} $\rm{\lambda}$$4026$. Each line was analyzed individually, with its corresponding spectral window automatically selected based on the respective synthetic profile.

The H$\alpha$ line is a fundamental marker for characterizing stellar winds, particularly in relation to mass-loss rates and the wind structure in massive stars. However, in O-type stars of luminosity class V and Vz, stellar winds are weak, which makes the contribution of H$\alpha$ less significant. Therefore, its absence in this analysis does not significantly impact the obtained results. Instead, the spectral characterization has focused on lines such as He I and He II, which are more relevant for these stars.\\

{\em{Effective Temperature ($T_\mathrm{eff}$), surface gravity ($\log {\mathrm{g}}$) and \mbox{helium abundance} ($\rm{Y_{He}}$):}} in our analysis, we adopted specific strategies to derive the stellar parameters. To do this, we selected initial values from previous estimates by \citep{massey2005physical}. Then, we calculated the chi-squared value by comparing the observed spectrum with the synthetic model at each point. During the iterative fitting process, the values of $T_\mathrm{eff}$, $\log {\mathrm{g}}$, $Y(\mathrm{He})$, among other relevant parameters, were adjusted. The goal was to find the set of parameters that allows the synthetic model to best fit the observational data.

For most of the stars analyzed, estimates for these parameters, along with their formal uncertainties, were derived. However, in some cases, due to data limitations or the inherent complexities of stellar characteristics, only upper or lower \mbox{limits} could be established for these parameters. It is important to note that the values of $\log {\mathrm{g}}$ and $T_\mathrm{eff}$ are not independent, as adjusting one influences the other. Therefore, IACOB-GBAT optimizes both parameters, along with others, during the fitting process to obtain a solution that is consistent with the spectroscopic observations.

Following previous \mbox{studies} \citep{herrero2002fundamental, ramirez2017vlt}, helium abundance was treated as a free parameter in this work.\\

{\em{Wind parameter ($\log {\mathrm{Q}}$):}} the wind strength parameter for O\,V and O\,Vz stars is generally undetermined due to their weak winds. Additionally, the lack of precise terminal \mbox{velocity} ($\mathrm{v_{\infty}}$) measurements introduces uncertainty when evaluating potential differences between O\,V and O\,Vz stars. Therefore, we focused on determining $\log {\mathrm{Q}}$ directly from the spectroscopic analysis. 

The spectroscopic determination of $\log {\mathrm{Q}}$ typically relies on the H$\alpha$ line, which is known to be highly sensitive to variations in mass-loss rate. However, in our case, the use of H$\alpha$ is severely limited due to strong nebular contamination in the NGC 346 region. This contamination significantly alters the observed profile, making it unsuitable for a reliable assessment of stellar wind properties in our sample.
Given this limitation, we based our analysis of $\log {\mathrm{Q}}$ on the \text{He\,II} $\rm{\lambda}$$4686$ line, which remains a valid diagnostic in this context. As discussed by \citet{sabin2014vlt},  the core flux of the \text{He\,II} $\rm{\lambda}$$4686$ line is sensitive to variations in $\log {\mathrm{Q}}$, making it  suitable for determining wind strength in O-type dwarfs.\\

{\em{Wind velocity exponent (\rm{$\beta$}) and microturbulence (\rm{$\xi_t$}):}} \mbox{according} to \cite{holgado2018iacob}, the IACOB-GBAT analysis cannot provide reliable constraints for the wind velocity exponent, especially within the considered range of values $(0.8-1.2)$. They also indicate that uncertainties in the $\log {\mathrm{Q}}$ parameter are reduced when $\beta$ is fixed. Therefore, we decided to set $\beta=1.0$, a value commonly used in O-type star studies. \\

Following recent studies \citep{mokiem2005spectral,ramirez2017vlt}, we treat microturbulence as a free parameter in our spectroscopic analysis. 
Determining microturbulence in O-type stars is particularly challenging due to the limited number of suitable metallic lines available for diagnostic purposes, in contrast to B-type stars, which offer a richer set of metal lines \citep{holgado2018iacob}. Although He lines provide some insight, their diagnostic power is limited. Holgado et al. also showed that certain diagnostic line cores are sensitive to microturbulence through saturation effects; however, this sensitivity is generally weak in O-type stars, limiting the precision of derived values.

Microturbulence affects the shape of spectral lines, especially metallic ones. High microturbulence leads to slight saturation in the line cores, resulting in broader and more squared or flattened profiles. This effect arises because microturbulence introduces non-thermal motions in the stellar atmosphere, increasing absorption at the line center and modifying the intensity distribution \citep{repolust2004stellar}. In contrast, low microturbulence produces more triangular-shaped profiles, as absorption is less influenced by turbulent motions.

\citet{sabin2014vlt} showed that the $\chi^2$ distribution for microturbulence is often degenerate, making it difficult to determine a unique value. In some cases, a slightly better fit is obtained with $\xi_t = 5\,\mathrm{km\,s^{-1}}$, but overall, microturbulence cannot be reliably constrained for O-type stars and 
is therefore commonly treated as a free parameter, 
an approach we adopt in this study.
While different values are tested during the fitting process, variations within the adopted range have minimal impact on the final stellar parameters \citep{villamariz2000fundamental}.

\section{Results}\label{ch:Results}

\subsection{NGC 346}

\subsubsection{Spatial and spectral type distributions}
\begin{figure}[ht!]
\centering
\includegraphics[width=9.0cm, height=7.5cm]{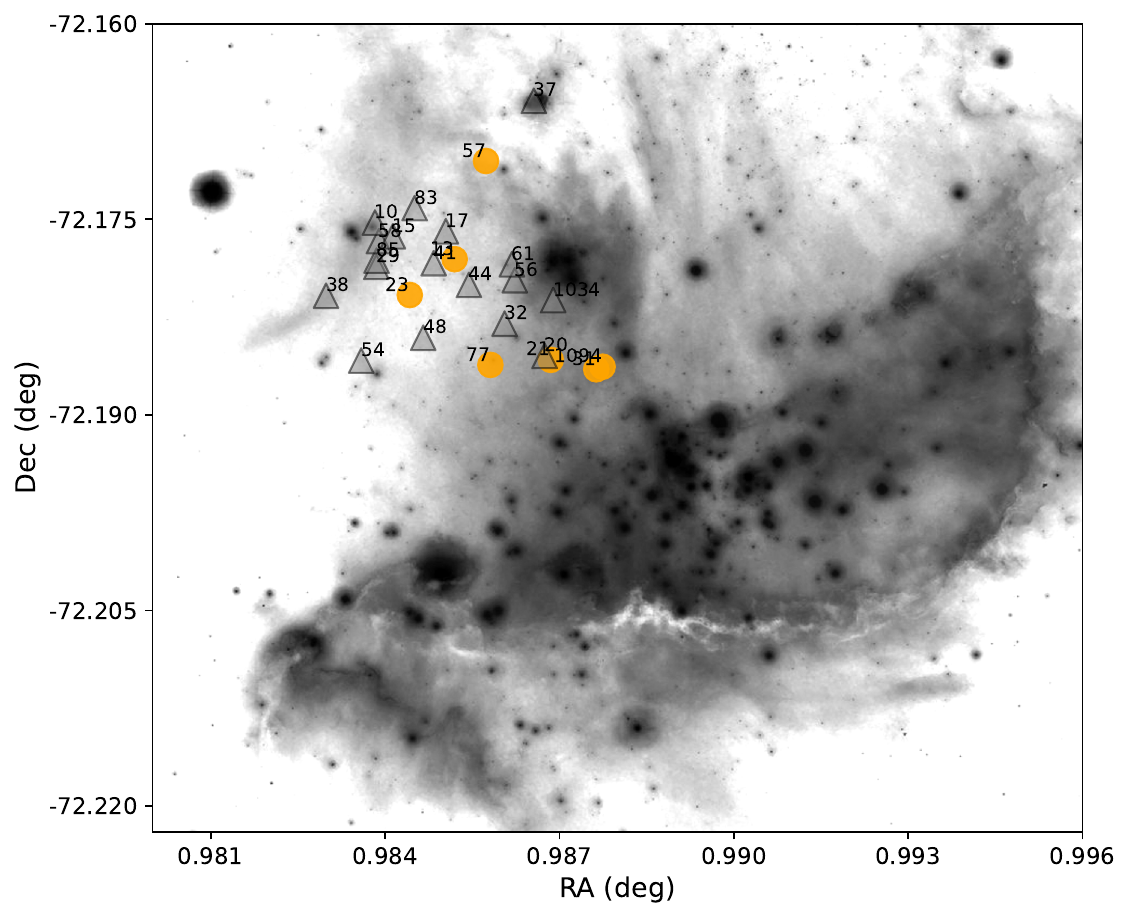}
\caption{Spatial distribution of O\,Vz stars (orange circles) and O\,V stars (gray triangles) analyzed in the NGC 346 region. Image Credit: NASA, ESA, CSA, N. Habel (JPL), P. Kavanagh (Maynooth University).}
\label{Figure:Figure_2}
\end{figure}

Table~\ref{Table:Table_3} provides detailed information on the updated classifications, equivalent widths, and their associated errors for the NGC 346 sample. Likewise, Fig.~\ref{Figure:Figure_2} illustrates the spatial distribution of O-type stars in the same sample. The luminosity subclasses are represented by gray triangles for O\,V stars and orange circles for O\,Vz stars. Normal dwarfs (O\,V) dominate the sample, significantly outnumbering their Vz counterparts. The O\,Vz stars are grouped into two distinct regions: one in the southern part and another across the nebula. In contrast, O\,V stars are more widely distributed along the periphery, with some overlapping with the southern O\,Vz group.

\begin{table*}[ht!]
\caption{Estimated equivalent widths (with their errors) of the spectral lines of He I and He II, along with the assigned parameter z to identify the spectral feature Vz in the analyzed stars in NGC~ 346.}
\label{Table:Table_3}
\centering
\begin{tabular}{lllcccccl}

\hline\hline
\noalign{\smallskip}
\text{Star} & \multicolumn{2}{c}{\small{He I $\lambda$4471}} & \multicolumn{2}{c}{\small{He II $\lambda$4542}} & \multicolumn{2}{c}{\small{He II $\lambda$4686}} & \text{z} & \text{Spectral Type} \\
            & \small{EW} & \small{$\sigma$} & \small{EW} & \small{$\sigma$} & \small{EW} & \small{$\sigma$} &           &                   \\
\hline
\noalign{\smallskip}
\text{NGC 346-10}&0.26 &$0.01$&0.72&$0.01$&0.73&$0.01$&1.02&  \text{O5.5\,V((f+))}\\
\text{NGC 346-13}&0.46 &$0.01$&0.42&$0.01$&0.73&$0.01$&1.58& \text{O8\,Vz}\\
\text{NGC 346-15}&0.46 &$0.01$&0.99&$0.01$&0.84&$0.01$&0.86& \text{O6.5\,V}\\
\text{NGC 346-17}&0.64 &$0.01$&0.20&$0.01$&0.44&$0.01$&0.69& \text{O9.2\,V}\\
\text{NGC 346-20}&0.42 &$0.01$&1.00&$0.01$&0.77&$0.01$&0.77& \text{O7\,V}\\
\text{NGC 346-21}&0.55 &$0.01$&0.48&$0.01$&0.62&$0.01$&1.10& \text{O7\,Vz}\\
\text{NGC 346-23}&0.60 &$0.01$&0.59&$0.01$&0.79&$ 0.01$&1.32& \text{O7.5\,Vz}\\
\text{NGC 346-29}&0.80 &$0.01$&0.78&$0.01$&0.53&$0.01$&0.66& \text{O9.5\,V}\\
\text{NGC 346-31}&0.26 &$0.01$&0.77&$0.01$&0.87&$0.01$&1.13& \text{O5.5\,Vz}\\
\text{NGC 346-32}&0.77 &$0.01$&0.41&$0.01$&0.55 &$0.01$&0.72& \text{O9.5\,V}\\
\text{NGC 346-37}&0.49 &$0.03$&1.03&$0.01$&0.69&$0.01$&0.67& \text{O8\,V}\\
\text{NGC 346-38}&0.74 &$0.01$&0.71&$0.02$&0.75 &$0.01$&1.01& \text{O9\,V}\\
\text{NGC 346-41}&0.80 &$0.01$&0.33&$0.01$&0.50&$0.01$&0.62& \text{O9.5\,V}\\
\text{NGC 346-44}&0.73 &$0.01$&1.14&$0.01$&0.85&$0.01$&0.75& \text{O8\,Vn}\\
\text{NGC 346-48}&0.63 &$0.01$&0.77&$0.01$&0.75&$0.01$&0.98& \text{O8\,V}\\
\text{NGC 346-54}&0.43 &$0.01$&1.01&$0.01$&0.96&$0.01$&0.95& \text{O6.5\,V}\\
\text{NGC 346-56}&0.70 &$0.01$&0.78&$0.01$&0.81&$0.01$&1.04& \text{O8\,V}\\
\text{NGC 346-57}&0.52 &$0.02$&0.78&$0.01$&0.93&$0.02$&1.19& \text{O7\,Vz}\\
\text{NGC 346-58}&0.86 &$0.01$&0.40&$0.01$&0.64&$0.01$&0.74& \text{O9.5\,V}\\
\text{NGC 346-61}&0.70 &$0.02$&0.36&$0.02$&0.65&$0.02$&0.93& \text{O9.5\,V}\\
\text{NGC 346-77}&0.59 &$0.01$&0.61&$0.02$&0.71&$0.01$&1.16& \text{O7.5\,Vz}\\
\text{NGC 346-83}&0.83 &$0.01$&0.27&$0.01$&0.48&$0.01$&0.57& \text{O9.5\,V}\\
\text{NGC 346-85}&0.77 &$0.01$&0.12&$0.01$&0.40&$0.01$&0.52& \text{O9.7\,V}\\
\text{NGC 346-1034}&0.77 &$0.01$&0.38&$0.01$&0.49&$0.01$&0.63& \text{O9\,V}\\
\text{NGC 346-1067}&0.30&$0.01$&0.44&$0.02$&0.70&$0.02$&1.51& \text{O8.5\,Vz}\\
\text{NGC 346-1094}&0.58&$0.02$&0.50&$0.01$&0.68&$0.01$&1.18& \text{O8\,Vz}\\
\text{NGC 346-1114}&0.69&$0.02$&0.30&$0.02$&0.58&$0.02$&0.84& \text{O9\,V}\\
    \text{NGC 346-1144}&0.46&$0.02$&0.28&$0.01$&0.47&$0.01$&1.01& \text{O9.5\,V}\\
\noalign{\smallskip}
\hline
\end{tabular}
\end{table*}

\begin{figure}[ht!]
\centering
\includegraphics[width=8cm, height=7cm]{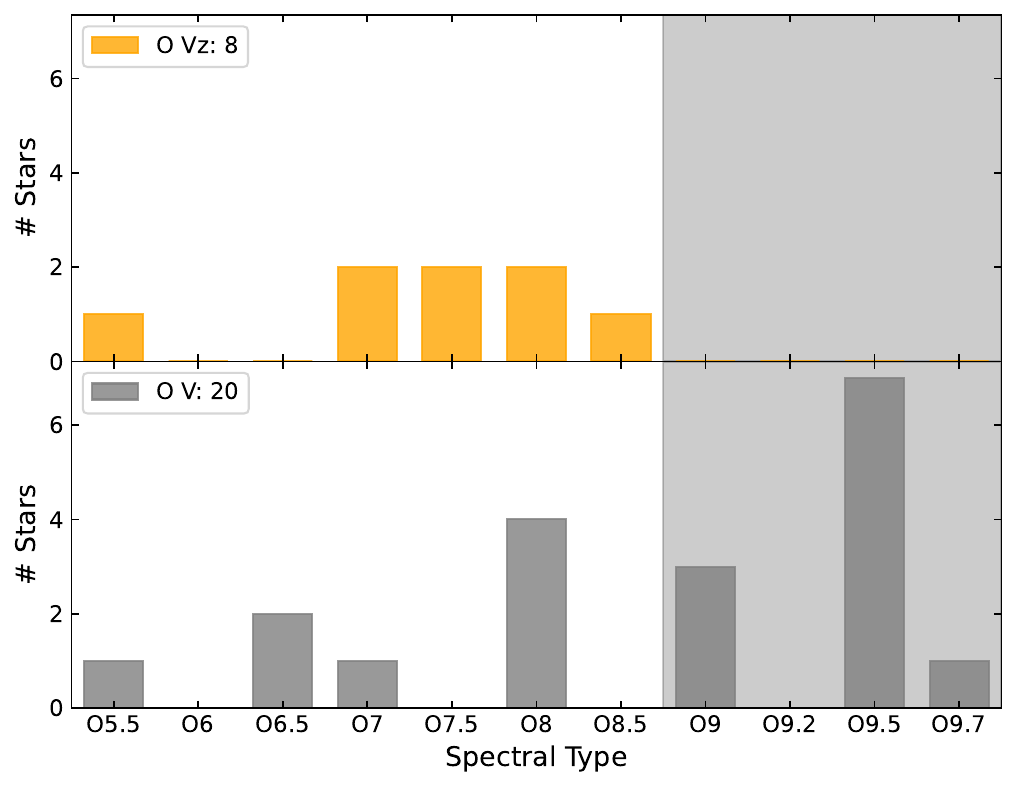}
\caption{Spectral distribution of O\,Vz (orange) and O\,V (gray) stars analyzed in the NGC 346 region. The gray dashed area represents the region defined by A16, where the Vz spectral feature is absent in stars with spectral later type than O8.5.}
\label{Figure:Figure_3}
\end{figure}

Figure.~\ref{Figure:Figure_3} shows the spectral type distribution of O-type stars in the NGC 346 sample. The upper panel shows the O\,Vz \mbox{population} (orange), while the lower panel represents the O\,V stars (gray). The normal dwarfs sample is dominated by mid- and late types, with a peak at spectral types O9 and O9.5 for the O\,V stars. In contrast, the O\,Vz stars are predominantly concentrated in intermediate spectral types between O5.5 and O8.5. The quantitative classification criterion applied in this study reveals that the O\,Vz stars constitute approximately $\sim$$29\%$ of the total sample, comprising eight objects. These stars are absent in later spectral types (> O8.5), consistent with prior findings. On the other hand, the O\,V stars, which represent $\sim$$71\%$ of the sample, are more evenly distributed across the spectral range but show a concentration in types later than O8.

\subsubsection{Distribution of the projected rotational velocities}
\begin{figure}[ht!]
\centering
\includegraphics[width=8cm, height=7cm]{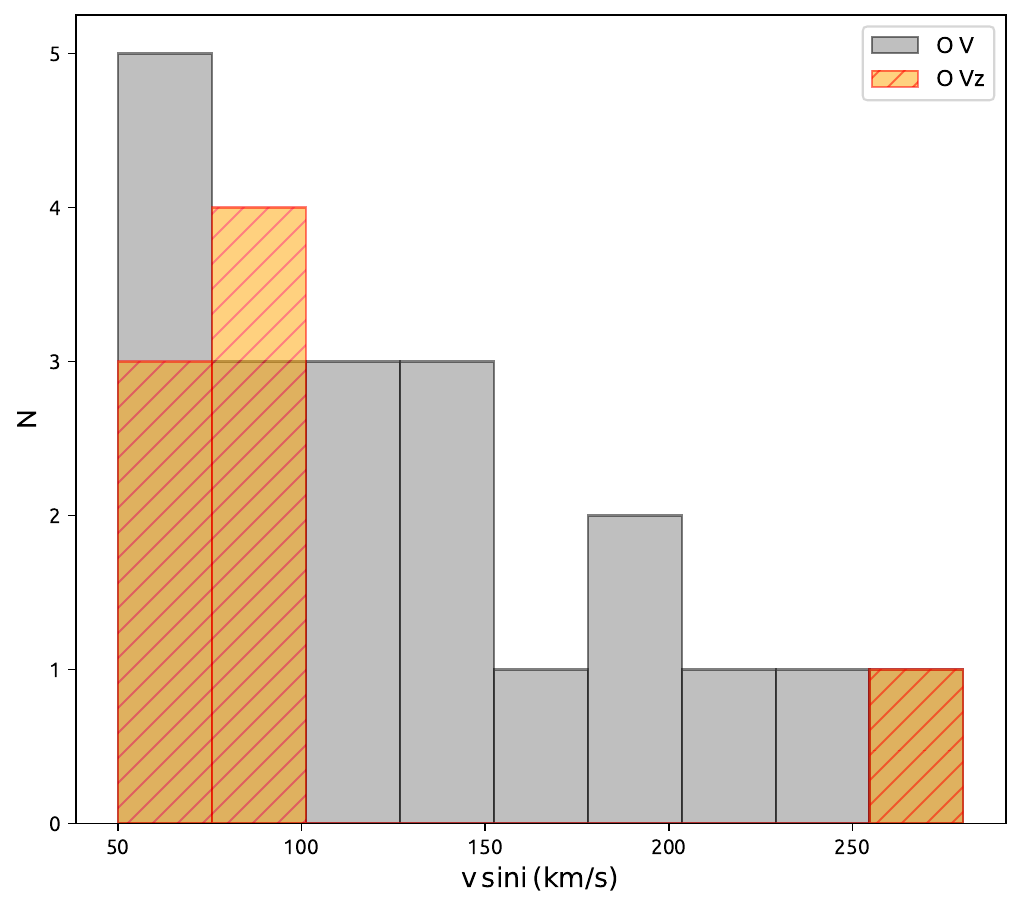}
\caption{Projected rotational velocity $v \, \mathrm{sin} \, i$ distribution of O\,Vz stars (orange diagonal lines) and O V stars (gray diagonal lines) analyzed in the NGC 346 region.}
\label{Figure:Figure_4}
\end{figure}

Figure.~\ref{Figure:Figure_4} shows the distribution of projected rotational velocity $v \, \mathrm{sin} \, i$ [$\rm{km\,s^{-1}}$] for O\,V (gray diagonal lines) and O\,Vz (orange diagonal lines) stars. As explained in Section.~\ref{ch:parameters}, for six O\,V and two O\,Vz stars where $v \, \mathrm{sin} \, i$ could not be reliably measured due to the resolution limit, a fixed value of $50$ $\rm{km\,s^{-1}}$ was adopted.

The O\,V population shows a wide range of $v \, \mathrm{sin} \, i$ values, with a clear concentration below $100$ $\rm{km\,s^{-1}}$, and a considerable fraction of stars extending toward higher rotational velocities (>$200$ $\rm{km\,s^{-1}}$). In contrast, the O\,Vz stars are predominantly concentrated at lower velocities ($v \, \mathrm{sin} \, i$ < $100$ $\rm{km\,s^{-1}}$), with only one object showing a higher value. While the overall distributions of O\,V and O\,Vz stars overlap significantly, the concentration of O\,Vz stars in the lower velocity range is noteworthy. However, the limited number of O\,Vz stars in the sample prevents a robust statistical comparison between the two populations. These \mbox{differences}, albeit tentative, could provide insights into the physical and evolutionary properties of these subclasses.\\

\subsubsection{Effective temperatures and surface gravities}

\begin{figure}[ht!]
\centering
\includegraphics[width=8.5cm, height=7cm]{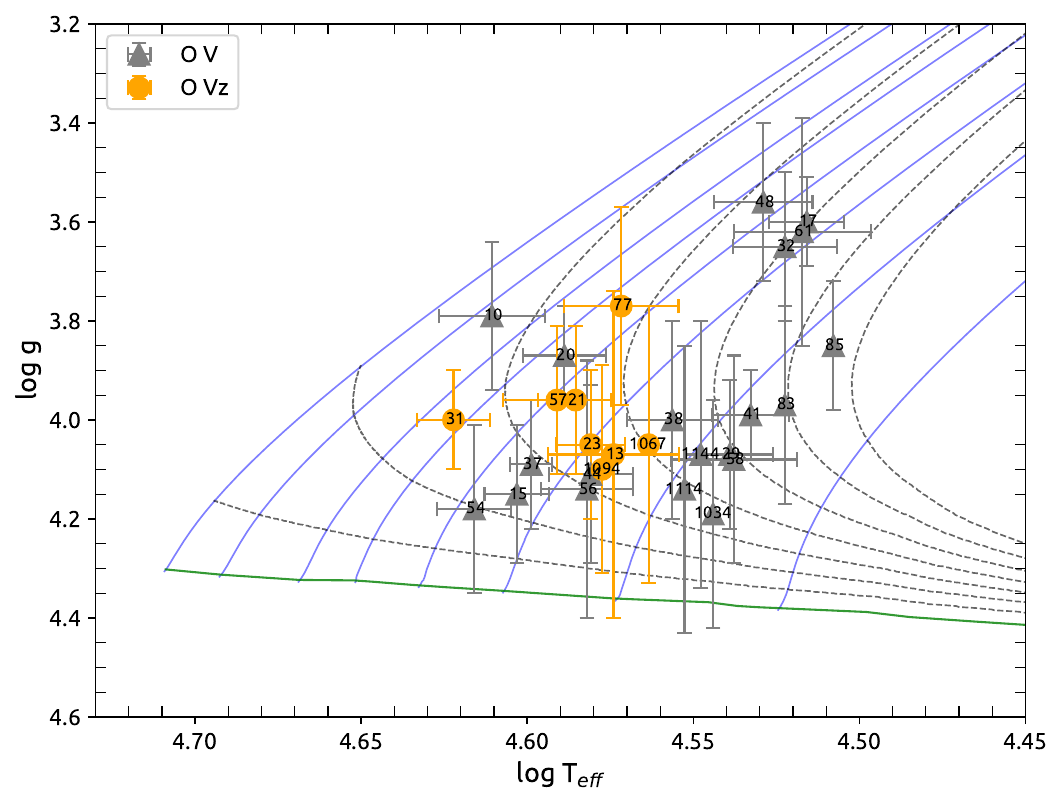}
\caption{Results of $\log {\mathrm{g}}$ vs $\log T_{\mathrm{eff}}$ for the O\,Vz and O\,V samples (orange circles and gray triangle symbols, respectively) in the NGC 346 region.}
\label{Figure:Figure_5}
\end{figure}

Figure.~\ref{Figure:Figure_5} shows the distribution of the analyzed stars in the $\log {\mathrm{g}}$ versus $\log T_{\mathrm{eff}}$ plane, plotting the ZAMS (green solid line), evolutionary tracks (blue solid lines), and isochrones (dashed lines) that cover ages up to 7 million years, based on the \mbox{models} of \cite{brott2011rotating}. These \mbox{models} were calculated for the characteristic metallicity of the SMC and assuming an initial rotational velocity of $180$ $\rm{km\,s^{-1}}$ \citep{dufton2019census}.   

As expected from their spectral-type distribution, O\,V stars show a broad dispersion in effective temperature, with \mbox{values} ranging from approximately $32$~\mbox{kK} to $40$~\mbox{kK}, with a peak frequency in $33$~\mbox{kK} and $36$~\mbox{kK}. In contrast, O\,Vz stars are \mbox{constrained} to a narrower range ($37$–$40$~\mbox{kK}) but overlap with O\,V stars in this interval.

Regarding surface gravities, O\,Vz stars are predominantly concentrated between $\log {\mathrm{g}}$ $\sim$ $3.7$~\mbox{dex} and $\log {\mathrm{g}}$ $\sim$ $4.1$~\mbox{dex}, clustering near $\log {\mathrm{g}}$ $\sim 4.0$~\mbox{dex}. O\,V stars, however, exhibit a broader range ($\log {\mathrm{g}}$ $\sim$ 3.6$-$ 4.2~\mbox{dex}) with concentrations at both ends of the distribution. 
The absence of O\,Vz stars in regions of lower surface gravity ($\log {\mathrm{g}}$$\sim$ $3.6$~\mbox{dex}) supports the hypothesis that these stars are less evolved and have not undergone significant expansion. However, the derived $\log {\mathrm{g}}$ values reveal that O\,Vz stars are not systematically closer to the ZAMS than O\,V stars, challenging the assumption that higher gravities consistently distinguish the Vz subclass. This finding aligns with previous analyses (SS14), highlighting the need for further investigation into the physical properties that drive the classification of O\,Vz stars.

\subsubsection{Helium abundance}

\begin{figure}[ht!]
\centering
\includegraphics[width=8cm, height=7cm]{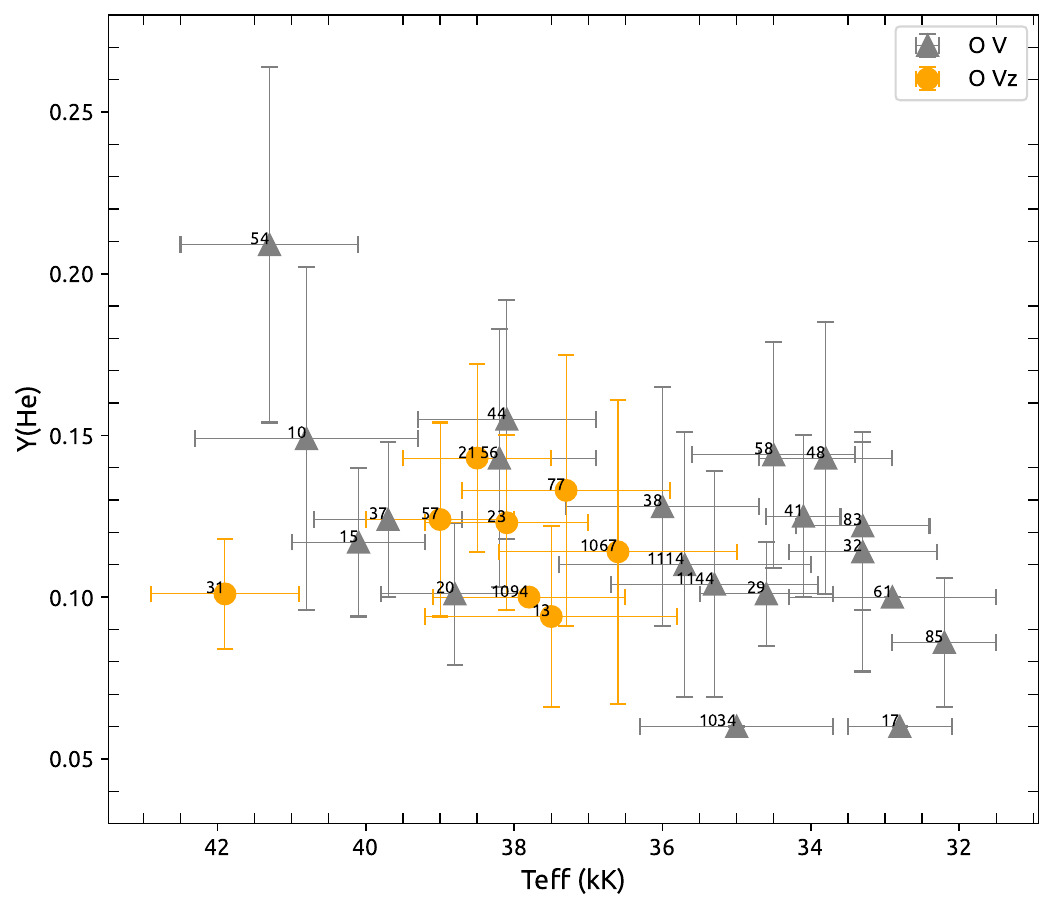}
\caption{$Y(\mathrm{He})$ vs $T_{\mathrm{eff}}$ of O\,Vz stars (orange circles) and O\,V stars (gray triangles) analyzed in the NGC 346 region.}
\label{Figure:Figure_6}
\end{figure}

Figure.~\ref{Figure:Figure_6} displays the helium abundance, $Y(\mathrm{He})$, as a function of the effective temperature, $T_{\mathrm{eff}}$, for the NGC 346 sample. No systematic difference in helium abundance is observed between O\,V and O\,Vz stars. Both subclasses exhibit a comparable dispersion in $Y(\mathrm{He})$, ranging from approximately $0.08$ to $0.14$, with significant variability within each spectral type. 

Three stars --NGC 346-17, NGC 346-1034 and \mbox{NGC 346-54}-- exhibit helium abundances that deviate significantly from theoretical predictions, with values $Y(\mathrm{He})$ = $0.06$, $0.06$, and $0.21$, \mbox{respectively}. These discrepancies may reflect complex physical conditions or highlight limitations in the modeling assumptions for these specific objects. Spectral fits assuming a fixed \mbox{$Y(\mathrm{He})$ = $0.10$}, consistent with the majority of the sample, proved unsatisfactory, emphasizing the need for further investigation.

\subsubsection{Wind-strength Q-parameter}
\begin{figure}[ht!]
\centering
\includegraphics[width=8cm, height=7cm]{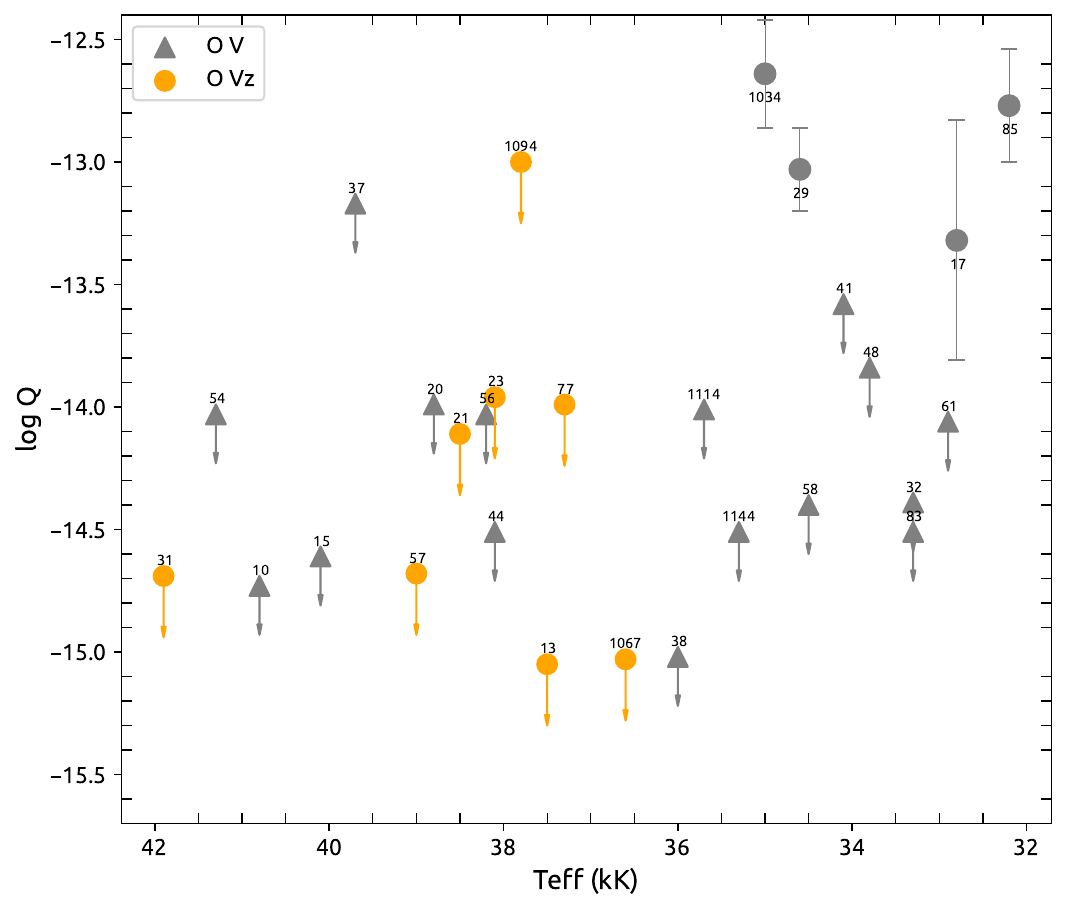}
\caption{Wind parameter $\log {\mathrm{Q}}$ vs $T_{\mathrm{eff}}$ of O\,Vz stars (orange circles) and O\,V stars (gray triangles) analyzed in the NGC 346 region.}
\label{Figure:Figure_7}
\end{figure}

To investigate the wind properties of the analyzed sample, Fig.~\ref{Figure:Figure_7} shows the derived $\log {\mathrm{Q}}$ values as a function of effective temperature. The downward arrows indicate upper limits for $\log {\mathrm{Q}}$, highlighting degeneracies at lower values and the impossibility of precise determinations in these cases.

O\,V and O\,Vz stars exhibit similar distributions in $\log {\mathrm{Q}}$  with no clear separation between the two groups. The most negative values of $\log {\mathrm{Q}}$, which correspond to the weakest stellar winds, do not show a direct association with either class. This suggests that the observed dispersion likely reflects the intrinsic \mbox{variability} of wind properties across the sample rather than systematic \mbox{differences} between O\,V and O\,Vz stars.

\subsubsection{The Hertzsprung-Russell (HR) diagram}

\begin{figure}[ht!]
\centering
\includegraphics[width=8cm, height=7cm]{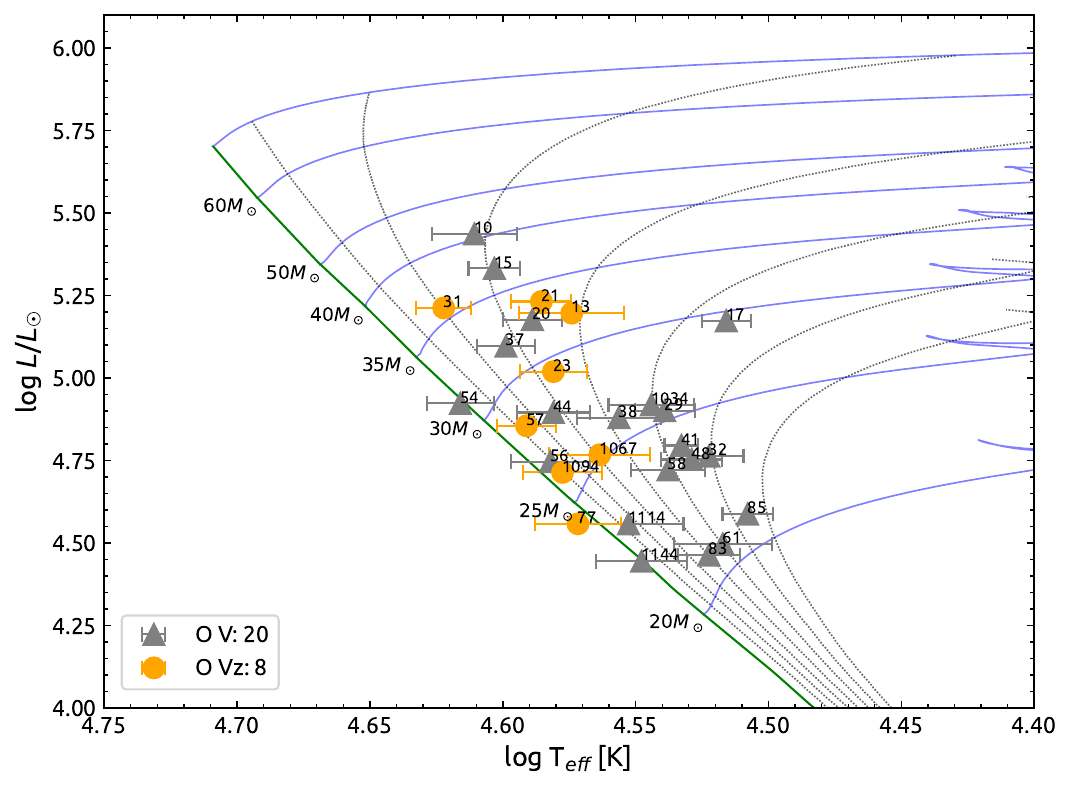}
\caption{Hertzsprung-Russell diagram for O\,Vz (orange circles) and O\,V (gray triangles) stars analyzed in the NGC 346 region.}
\label{Figure:Figure_8}
\end{figure}

In Fig.~\ref{Figure:Figure_8} we present the Hertzsprung-Russell (HR) diagram for the NGC 346 region. The ZAMS (solid green line), \mbox{evolutionary} tracks (solid blue lines), and isochrones (dotted lines) are the same as those shown in Fig.~\ref{Figure:Figure_5}. The luminosities were determined \mbox{following} the methodology proposed by \citet{hunter2007vlt}, \mbox{using} V-band magnitudes estimated by \citet{massey1989stellar}. Due to the low extinction toward the Magellanic Clouds, a single reddening value for NGC 346 was adopted, as suggested by \citet{dufton2019census}. The reddening law by \citet{bouchet1985visible} was used, expressed \(\rm{A_V = 2.72 E(B-V)}\), with a constant \mbox{reddening} value of \(\rm{E(B-V)=0.09}\) \citep{massey1989stellar}. Bolometric corrections from \citet{vacca1996lyman} were applied, assuming a distance \mbox{modulus} of $18.91$ \citep{cignoni2010history}.

The diagram reveals a sparsely populated region above the upper ZAMS, with no stars younger than $1$~\mbox{Myr} and hotter than $40$~\mbox{kK}, suggesting a dearth of very massive stars. These findings are consistent with the recent analysis by \citet{rickard2022stellar}, who also noted a deficiency of O-type stars on the ZAMS and in the upper region of the HR diagram, based on a sample of 18 O-type stars with luminosity class V and 3 with luminosity classes II–III. Regarding ages, most stars in the 28$-$35~\mbox{M$_{\odot}$} mass range exhibit values between 1 and 4 Myr, with minimal dispersion, while stars with masses below $25$ $M_{\odot}$ show a broader age range, spanning from $1$ to $6$ Myr. Notably, the ages of O\,Vz stars and O\,V stars show no systematic differences, once again calling into question the potential link between the Vz classification and a younger evolutionary stage.

\subsection{\textbf{30~Doradus revisited}}

\subsubsection{Spectral reclassification}

\begin{figure*}[ht!]
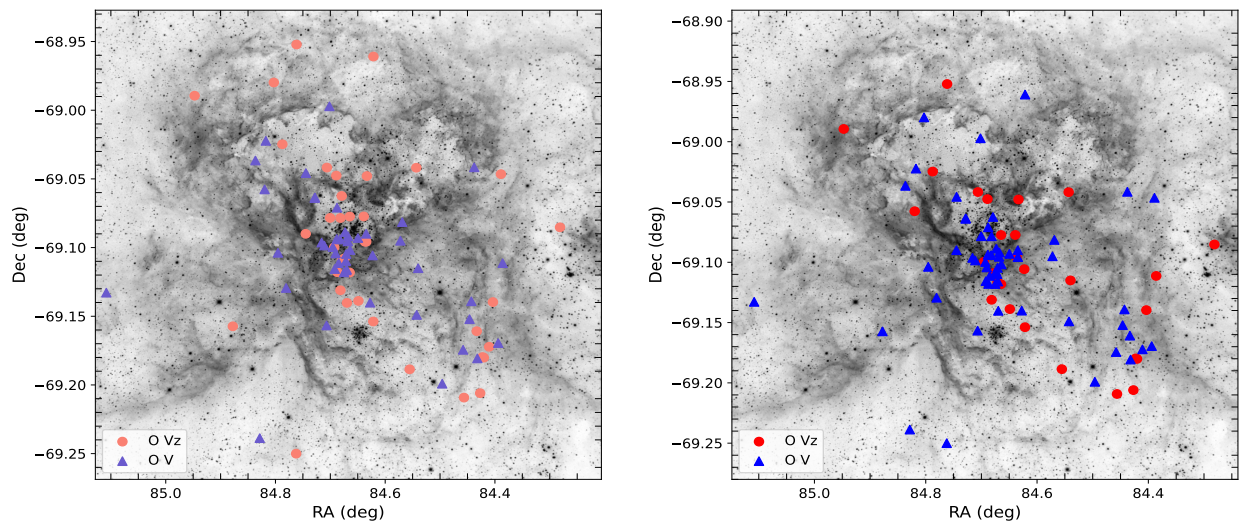

\centering
\includegraphics[width=8cm, height=7cm]{Figures/Figure_9a.png}
\hspace{0.2cm}
\includegraphics[width=8cm, height=7cm]{Figures/Figure_9b.png}
\caption{Spatial distribution of O\,Vz (red circles) and O\,V (blue triangles) stars analyzed in the 30~Doradus region by \citet{sabin2014vlt} (left panel) and this work (right panel).}
\label{Figure:Figure_9}
\end{figure*}

The O\,V and O\,Vz stars in 30~Doradus were reclassified using the quantitative criterion proposed by A16, which is more restrictive than the purely morphological approach employed by SS14. The quantitative method was designed to minimize the inclusion of marginal cases and provide a more robust classification based on the equivalent widths of specific helium lines. Table~\ref{Table:Table_5} provides information on the equivalent widths and their errors for this region.

Applying this updated criterion resulted in a slightly different distribution of O\,V and O\,Vz stars compared to SS14. Specifically, our results indicate a reduction of approximately $12\%$ in the number of O\,Vz stars, with $28$ stars now classified as O\,Vz (compared to $38$ in SS14) and $56$ as O\,V (compared to 46 in SS14). O\,Vz stars represent $33\%$ of the total sample. Despite this quantitative difference, the overall trends remain qualitatively consistent: O\,Vz stars are less numerous than O\,V stars and are distributed throughout the region, with no evident spatial differences between the two groups, as shown in Fig.~\ref{Figure:Figure_9}

\begin{figure*}[ht!]
\centering
\includegraphics[width=8cm, height=6.5cm]{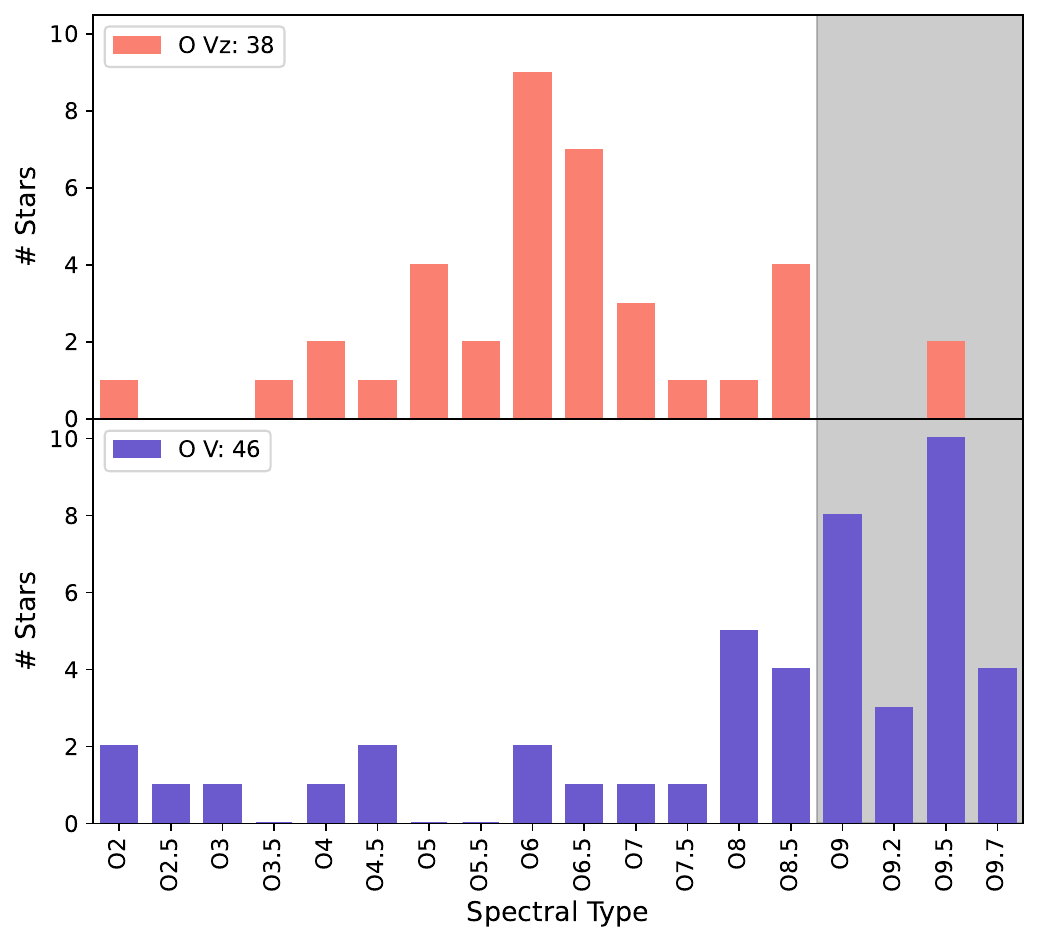}
\hspace{0.3cm}
\includegraphics[width=8cm, height=6.5cm]{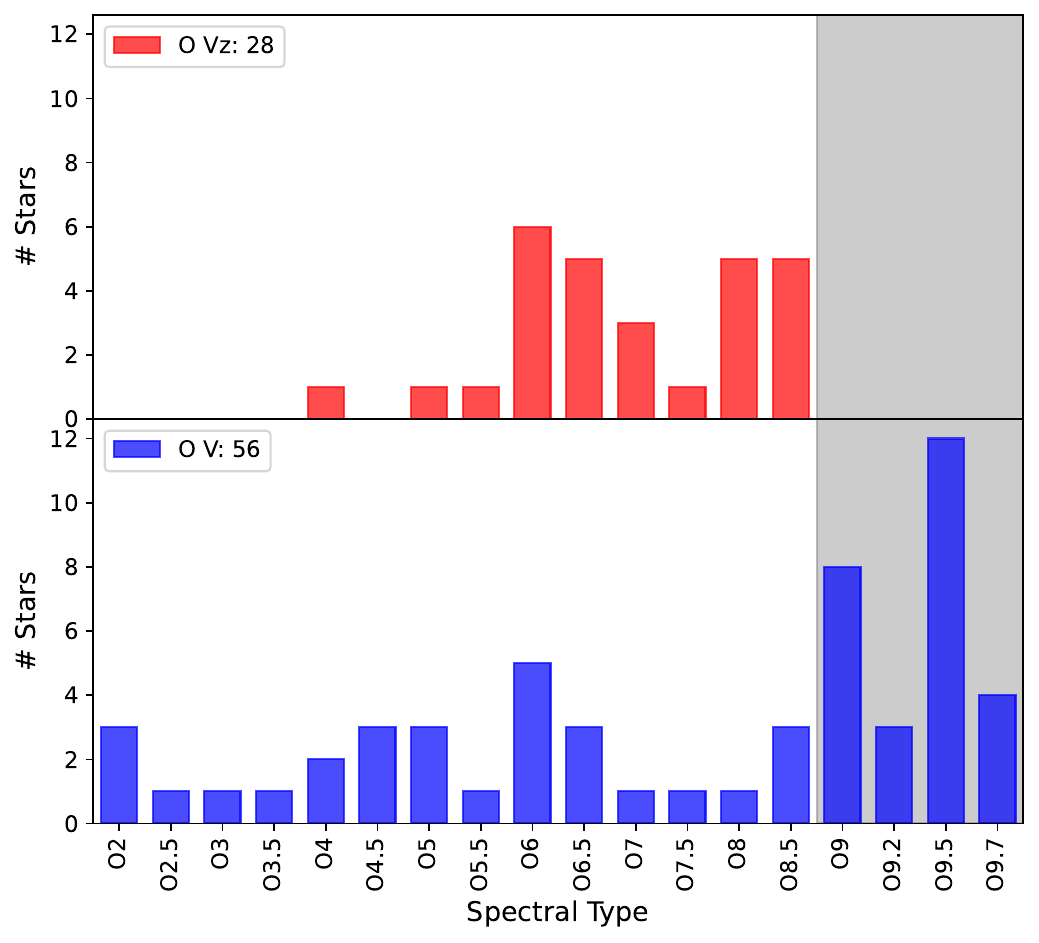}
\caption{Spectral distribution of O\,Vz (red) and O\,V (blue) stars analyzed in the 30~Doradus region by \citet{sabin2014vlt} (left panel) and this study (right panel). The gray dashed area corresponds to the one defined by A16, as shown in Fig.~\ref{Figure:Figure_3}.}
\label{Figure:Figure_10}
\end{figure*}

In Fig.~\ref{Figure:Figure_10}, the spectral type distributions of O\,V and O\,Vz stars are compared. Our findings confirm that O\,Vz stars are most common in intermediate spectral types (O5–O7), while O\,V stars dominate later spectral types. The application of the quantitative classification criterion results in a spectral distribution that is more astrophysically consistent: O\,Vz stars are fewer in number than O\,V stars, as expected from the shorter timescales associated with the Vz phenomenon, and no O\,Vz stars are found at late spectral types, in agreement with theoretical predictions.

\subsubsection{Comparison of the Hertzsprung-Russell diagrams}
\begin{figure*}[t]
\centering
\includegraphics[width=8cm, height=6.5cm]{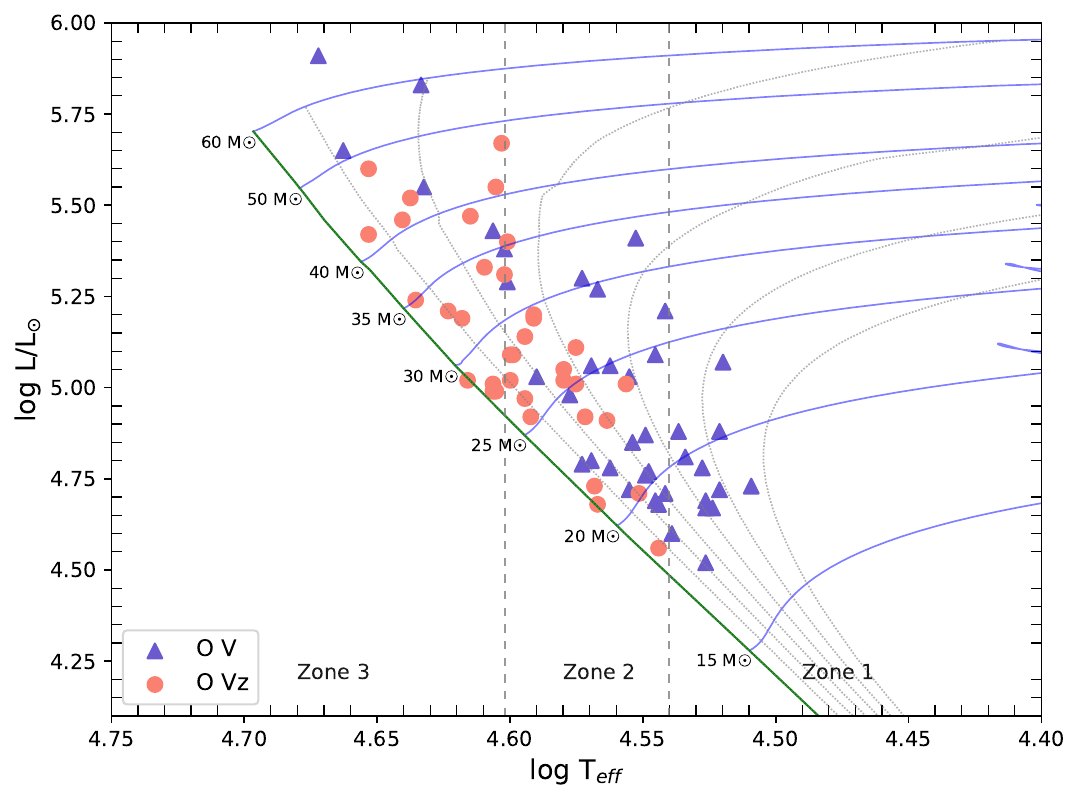}
\hspace{0.3cm} 
\includegraphics[width=8cm, height=6.5cm]{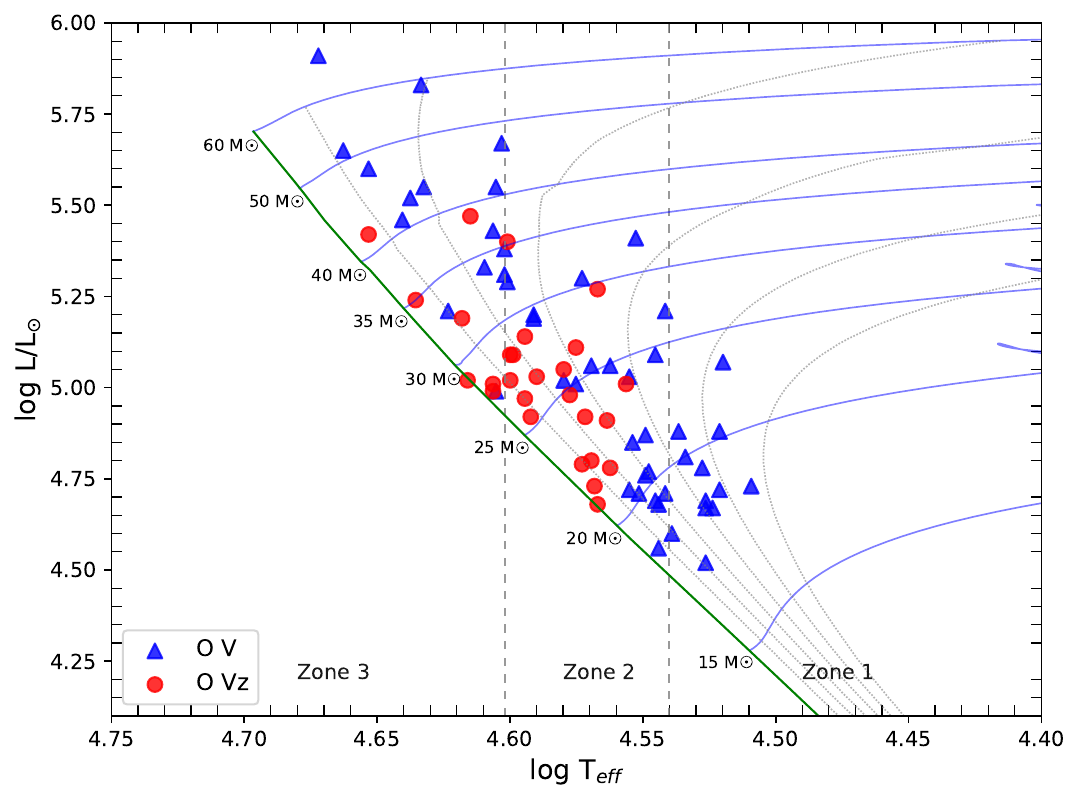}
\caption{Hertzsprung-Russell diagram for the O\,Vz and O\,V samples in 30~Doradus (represented by red circles and blue triangle symbols, respectively), analyzed by \citet{sabin2014vlt} (left panel) and in this study (right panel).  
Zone 1: $T_{\mathrm{eff}}$ $\leq$ 35\,000 K.
Zone 2: 35\,000 K $\leq$ $T_{\mathrm{eff}}$ $\leq$ 40\,000 K.
Zone 3: $T_{\mathrm{eff}}$ $\geq$ 40\,000 K).}
\label{Figure:Figure_11}
\end{figure*}

Figure.~\ref{Figure:Figure_11} presents two HR diagrams for 30~Doradus: the left panel shows the results based on the morphological \mbox{classifications} of SS14, while the right panel depicts a \mbox{reconstruction} \mbox{using} the updated spectral classifications derived in this study. Both diagrams reveal similar overall distributions. O\,Vz stars (red circles) are concentrated at higher effective temperatures \mbox{($T_{\mathrm{eff}}$ $\geq$ 35\,000 K)} and, in general, are located closer to ZAMS. In contrast, O\,V stars (blue triangles) occupy a broader range in the HR diagram, extending toward cooler ($T_{\mathrm{eff}}$ $\leq$ 35\,000 K) and more evolved regions. In both cases, a transition from O\,Vz to O\,V is observed as stars begin to develop stronger stellar winds, consistent with the evolutionary models of massive stars.

However, differences are apparent in the number of O\,Vz stars found at advanced evolutionary stages. The SS14 classifications include $16$ O\,Vz stars with ages exceeding 2 Myr (see Table~4 from SS14), whereas the stricter classification criteria employed in this study result in only 10 such stars. This aligns more closely with the hypothesis that O\,Vz stars represent a younger population within the O\,V class.

In any case, while the refined classification reduces the \mbox{presence} of O\,Vz stars with advanced ages in about a $37\%$, the Vz phenomenon still appears to persist for a longer duration in 30~Doradus compared to the Galaxy. As suggested by SS14, this extended persistence may be linked to the lower metallicity of the LMC. Following this reasoning, O\,Vz stars with ages \mbox{exceeding} 2 Myr could be expected to be more prevalent in the even lower-metallicity environment of the SMC.

\section{Discussion}\label{ch:Discussion}

In this study we investigated the O-type stars of luminosity classes V and Vz within the star-forming region NGC~346 in the SMC. Our analysis focused on comparing the physical and \mbox{evolutionary} properties of these two subclasses, drawing parallels with \mbox{previous} work in the 30~Doradus region (SS14). Below, we discuss our main findings and their implications.\\

{\em Proportion of O\,Vz stars in NGC 346.} Out of a sample of $28$ O-type stars of class V in NGC~346, 8 were identified as \mbox{belonging} to the Vz subclass, corresponding to $29\%$ of the sample. This proportion is consistent with the 33\% found in 30~Doradus after the spectral reclassification based on quantitative criteria. In the Milky Way, the systematic investigation of the Galactic O\,Vz stars conducted by \citep{arias2016spectral} identified $78$ O\,Vz stars out of a total of $197$ class V objects with spectral types between O3 and O8.5, corresponding to $39\%$ of the sample. However, comparing the Milky Way as a whole with the specific star-forming regions of the Magellanic Clouds introduces complications, given the heterogeneity of environments within the Galaxy. A more appropriate comparison would involve regions with similar \mbox{characteristics}, such as the massive star-forming cluster NGC~3603.

NGC~3603 hosts a rich population of O-type stars, \mbox{making} it an excellent candidate for such comparative studies. \mbox{Although} some spectroscopic studies of the massive stellar population in NGC~3603 exist \citep[e.g.,][]{melena2008massive}, none have \mbox{specifically} focused on characterizing the O-type class V stars, or \mbox{investigating} the presence of Vz stars. This represents an \mbox{opportunity} to explore the physical and evolutionary properties of these stars in a Galactic environment that shares similarities with 30~Doradus and NGC ~346. We aim to address such comparison in a forthcoming study.

\vspace{0.1cm}
{\em Comparison between O\,V and O\,Vz stars in NGC~346.}
The quantitative spectroscopic analysis of the NGC~346 sample reveals some differences between O\,V and O\,Vz stars, although these are insufficient to establish clear distinctions in terms of evolutionary status. Notably, no systematic differences are observed in the wind strength parameter $\log {\mathrm{Q}}$, which would be expected from the stronger $\rm{He \,II \,\lambda4686}$ absorption indicative of a potentially lower mass-loss rate. Similarly, no significant \mbox{differences} in helium abundance are observed between the two subclasses. Finally, O\,Vz stars do not exhibit systematically lower luminosities or higher gravities, which would be expected if they were exclusively younger objects nearer to the ZAMS (see the HR diagram in Fig. \ref{Figure:Figure_8}). These findings are consistent with the results of SS14 for 30~Doradus, supporting the hypothesis that the Vz classification is not a direct evolutionary status indicator, but arises from specific combinations of physical parameters.

The lack of clear evolutionary distinctions between O\,V and O\,Vz stars confirms that the Vz phenomenon results from a complex interplay of physical parameters, including effective \mbox{temperature}, surface gravity, the wind strength parameter and the projected rotational velocity. As demonstrated by SS14, variations in these four parameters affect the relative intensities of the He lines relevant to the Vz classification. As a result, stars classified as Vz may not necessarily represent a homogeneous or distinct group but rather a subset of O-type stars that happen to fall within a specific parameter space at the time of observation.

\vspace{0.1cm}
{\em Metallicity and the extended Vz phase.} SS14 proposed an explanation for the large number of O\,Vz stars found in 30~Doradus and their unexpected presence at significant distances from the ZAMS. They hypothesized that the low metal content of the LMC implies weaker stellar winds, preventing most O dwarfs, even relatively evolved, from breaking the Vz characteristic. In this context, fewer Vz stars would be expected away from the ZAMS in the Milky Way, where higher metallicity strengthens stellar winds and accelerates the loss of the Vz characteristic. \mbox{Extending} this reasoning to the SMC, with its even lower metallicity, a greater proportion of O\,Vz stars with advanced ages might be expected.

Our findings suggest a potential correlation between \mbox{metallicity} and the duration of the Vz phase. In NGC~346, where the metallicity is approximately 0.4 times that of 30~Doradus, $50\%$ of the O\,Vz stars have ages exceeding $2$ Myr, and $38\%$ are older than $3$ Myr. In contrast, only $11\%$ of O\,Vz stars in 30~Doradus surpass $3$ Myr. For comparison, the Galactic sample studied by \cite{holgado2020iacob} includes only one O\,Vz star older than $3$ Myr out of a total of $32$ stars. Although the size of our sample is very limited, this trend supports the hypothesis that lower metallicity extends the Vz phase by weakening stellar winds. The importance of stellar wind strength in shaping the observable properties of O stars is thus reinforced.

\vspace{0.1cm}
{\em On the absence of very massive stars near the ZAMS.}
The absence of very massive stars near the ZAMS is a topic of active discussion in stellar astrophysics. In an extensive study of Galactic O-type stars, \cite{holgado2020iacob} identified a dearth of ZAMS objects in the mass range of approximately 30$-$70 $M_{\odot}$. The authors demonstrated that this gap persists even after \mbox{accounting} for potential observational biases, such as extinction by the parental cloud. Similar findings have been reported in the Magellanic Clouds. For instance, \cite{schootemeijer2020dearth} used a combination of spectral types from catalogs and Gaia magnitudes to report a dearth of young and bright massive stars in the SMC. In 30~Doradus, \cite{schneider2018vlt} identified a sparsely occupied region in the HR diagram between the ZAMS, the 1~Myr isochrone, and the 30~$M_{\odot}$ stellar track among the OB population from the VFTS survey. This fact is also apparent in Fig.~\ref{Figure:Figure_11}.

In our study of NGC 346, no stars within the 1~Myr isochrone are observed above the 25~$M_{\odot}$ evolutionary track in the HR diagram, except for star NGC~346-54 (see Fig. \ref{Figure:Figure_8}). This result aligns with the trends reported for other regions but must be interpreted with caution given the small and likely incomplete sample \mbox{analyzed} here. The challenges posed by distance, \mbox{crowding}, and extinction in NGC~346 may contribute to the observed gap, highlighting the need for more comprehensive studies with larger datasets.

\section{Summary and Outlook}\label{ch:Summary}

In this study, we conducted a detailed spectroscopic analysis of O\,V and O\,Vz stars in NGC~346, the most O-star-rich star-forming region in the SMC. O\,Vz stars are particularly intriguing, as they have been hypothesized to represent a less evolved subgroup among the youngest, optically visible, massive stars. Using intermediate-resolution spectra obtained with the MagE spectrograph, complemented by public FLAMES data, we updated the spectral classifications of 37 stars and determined their physical parameters.

Our results indicate that the differences exhibited by O\,Vz stars in NGC~346 are insufficient to establish them as \mbox{systematically} younger or closer to the ZAMS compared to O\,V stars, consistent with previous findings in 30~Doradus. We confirm that the Vz classification is not a reliable indicator of evolutionary age but instead appears to result from a specific combination of stellar parameters. Nonetheless, the role of wind properties influenced by low-metallicity environments could be particularly significant. In this context, our findings suggest that the duration of the Vz phase may be correlated with metallicity, with stars in the SMC exhibiting this characteristic for longer periods due to their weaker stellar winds.

As we have mentioned throughout the Discussion Section, the primary limitation of this study is the small sample size. The \mbox{challenges} posed by NGC~346 — such as its distance and crowding — complicate the identification, classification, and detailed spectroscopic analysis of its stellar population. Another limitation is the relatively low resolution of the MagE spectra, which restricts the precision with which we can derive key stellar parameters. These limitations constrain the significance of our analyisys, highlighting the need for more comprehensive datasets.

Future surveys, such as those conducted under the \mbox{ULLYSES} program \citep{roman2020ultraviolet}, hold great promise for addressing these challenges. A key component is the \mbox{X-Shooting ULLYSES} subprogram \citep{vink2023x}, which uses the \mbox{X-Shooter} instrument on the VLT to obtain optical to near-infrared spectra of the ULLYSES sample. These data are particularly suitable for the detailed characterization of massive stars. 
While our study represents an initial step toward understanding the role of metallicity in the O\,Vz phenomenon, \mbox{X-Shooting ULLYSES} observations of NGC 346 will offer a broader perspective on these stars, helping to refine and expand our conclusions.

\begin{acknowledgements}
We thank the referee for their constructive comments, which helped improve the clarity of this work. This research is part of the Ph.D. thesis of L. Arango and is supported by the Agencia Nacional de Investigación y \mbox{Desarrollo} (ANID) through the Chilean National Doctoral Fellowship No.~21221152. We also acknowledge financial support from the Chilean Astronomical Society (SOCHIAS). J. I. A acknowledges the financial support from the Dirección de Investigación y Desarrollo de la Universidad de La Serena (ULS), through the project PR2324063. G. H acknowledges support from the Spanish \mbox{Ministry} of Science and Innovation and Universities (MICIU) through the Spanish State
Research Agency (AEI) through grants PID2021-122397NB-C21, PID2022-136640NB-C22, 10.13039/501100011033, and the Severo Ochoa Program 2020-2023 (CEX2019-000920-S).
\end{acknowledgements}

\bibliographystyle{aa} 
 \bibliography{Bibliografia}

\begin{appendix} 

\section{Spectral classification} \label{ch:clasificacion}
The determination of spectral types and luminosity classes for the stars in NGC 346 was carried out using a comparative approach. This involved a qualitative assessment in which the observed spectra were compared with the standards from the Galactic O-Star Spectroscopic Survey \citep[GOSSS,][]{sota2011galactic, sota2014galactic}, applying the criteria established in the catalog. Despite the limitations imposed by the spectral resolution of the instrument, the visual method proved to be suitable even for spectra with low to \mbox{moderate} S/N, allowing for a reliable classification in most cases \citep{melena2008massive}.

It is important to highlight that the resolving power of the stellar sample in NGC~346 ($\rm{R} \sim 4100$) is larger than that GOSSS classification atlas ($\rm{R} \sim 2500$). Consequently, the resolution of the spectra in the Magellanic Cloud region was degraded using the IRAF environment to properly interpret the spectral characteristics and apply the standard spectral classification for O-type stars, as outlined in the atlas. The degradation was carried out \mbox{considering} that the internal spectral resolution per pixel of the GOSSS survey corresponds to $\rm{0.7956}$ $\AA/\rm{pixel}$. 

\begin{figure}[ht!]
\includegraphics[width=8.5cm, height=9cm]{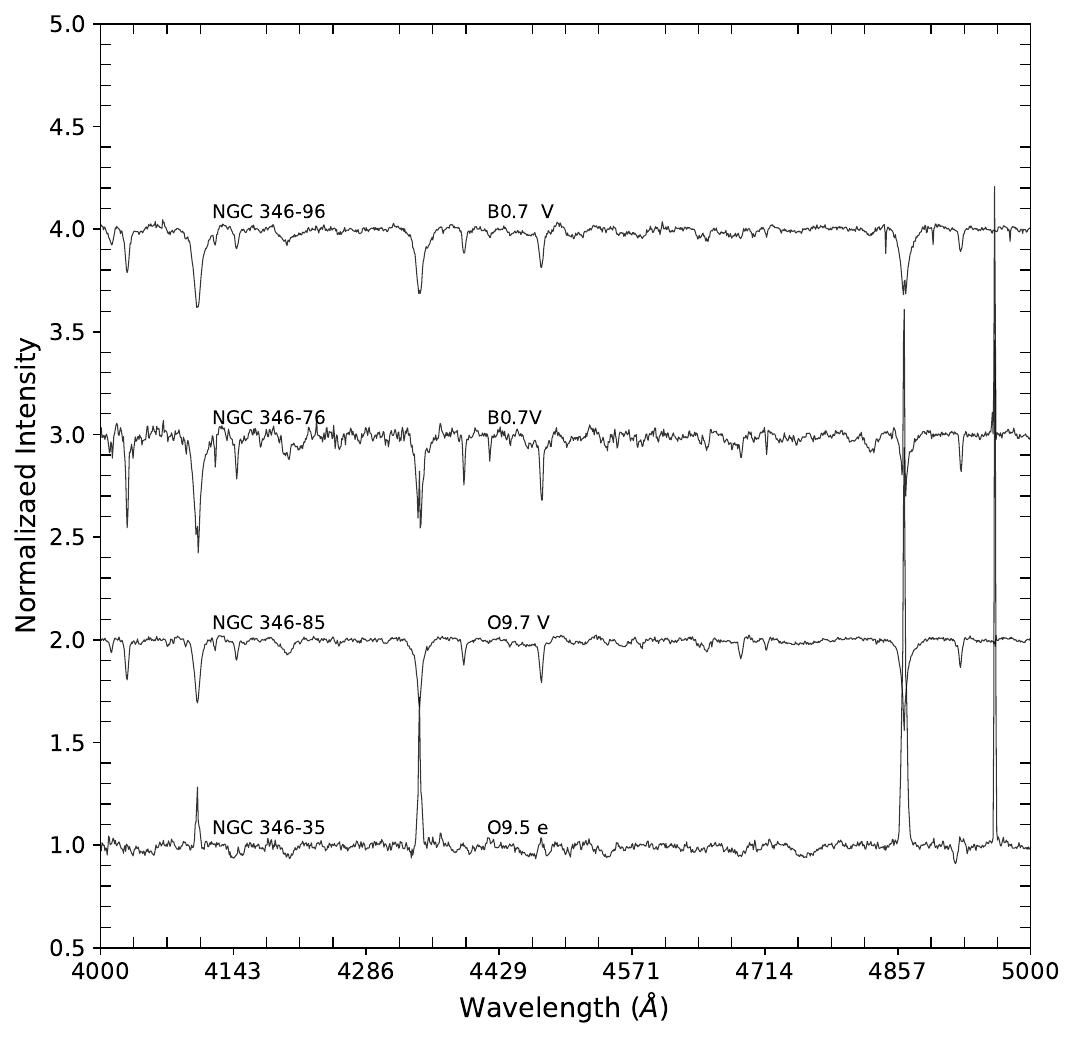}
\caption{Normalized spectra of newly classified dwarfs in NGC 346.}
\label{Figure:Figure_13}
\end{figure}

We have identified five previously unclassified stars in the cluster (see Fig.~\ref{Figure:Figure_13}). The new spectral types we have assigned are listed in Table~\ref{Table:Table_7}. 

\begin{table}[ht!]
\caption[]{NGC 346 stars with newly classified spectral types.}
\label{Table:Table_7}
$$
 \begin{tabular}{lr}
    \hline \hline
    \noalign{\smallskip}
\text{Star}  & \text{Spectral type} \\
\noalign{\smallskip}
\hline
\noalign{\smallskip}
\text{NGC 346-35}  & \text{O9.5\,e}\\
\text{NGC 346-85}  & \text{O9.7\,V}\\
\text{NGC 346-76}  & \text{B0.7\,V}\\
\text{NGC 346-96}  & \text{B0.7\,V} \\
 \noalign{\smallskip}
\hline
\end{tabular}
 $$
 \tablefoot{The stars NGC 346-35, NGC 346-76, and NGC 346-96 exhibited later spectral types within the NGC 346 sample. Since our analysis focuses exclusively on the O\,V luminosity class, these three stars were excluded from the initial studies. }
   \end{table}

\onecolumn
\section{Tables} \label{ch:Tables}
\begin{table}[ht!]
\caption{NGC 346 stars.}
\label{Table:Table_4}
\begin{tabular}{lcccclll}
\hline\hline
\noalign{\smallskip}
Star  & \text{Identifier} & \text{RA (J2000)} & \text{Dec (J2000)} &\text{V} &\text{Literature Spectral Type}  &\text{Comments} \\
\noalign{\smallskip}
\hline
\noalign{\smallskip}
NGC 346-06\tablefootmark{(1)}&\text{MPG} 342&00:59:00.01&-72:10:37.7&13.66&\text{O5-6\,V((f))\,(1-7)}&\text{Binary(SB3)\,(7)}\\
NGC 346-09\tablefootmark{(1)}&\text{MPG} 324&00:58:57.36&-72:10:33.4&14.02&\text{O4\,V((f+))\,(3)}& Binary\,(4)\\
NGC 346-10 &\text{MPG} 368&00:59:01.77&-72:10:30.9&14.18
&\text{O5.5\,V((f+))\,(1);}\\
& & & & & \text{O4-5\,V((f))\,(2);}\\
& & & & & \text{O6\,V((f))\,(7)}\\
NGC 346-13&\text{MPG} 487&00:59:06.71&-72:10:41.0&14.53
&\text{O8\,V\,(1)}\\
NGC 346-15&\text{MPG} 396&00:59:02.88&-72:10:34.7&14.39
&\text{O7\,V\,(1)}\\
NGC 346-16 \tablefootmark{(1)}&\text{MPG} 470&00:59:05.94&-72:10:33.6&13.78
&\text{O8.5\,III\,(1); O8\,V\,(8)}&\text{Eclipsing Binary\,(8)}\\
NGC 346-17&\text{MPG} 476&00:59:06.16&-72:10:33.3&14.19
&\text{O6\,V\,(8)}&\\
NGC 346-20&\text{MPG} 602&00:59:12.28&-72:11:07.7&14.68
&\text{O7\,V\,(1); O6.5\,V((f))\,(7)}\\
NGC 346-21&\text{MPG} 615&00:59:12.66&-72:11:08.8&14.52
&\text{O8\,V\,(1); O7\,V\,(7)}\\
NGC 346-23&\text{MPG} 417&00:59:03.93&-72:10:50.9&15.02
&\text{O7.5\,V((f))\,(1-7);}\\
& & & & & \text{O7.5\,Vz\,(8)}\\
NGC 346-26\tablefootmark{(1)}&\text{MPG} 495&00:59:07.29&-72:10:25.1&15.13
&\text{O8\,V\,(1); O9\,V\,(7)}&\text{Binary\,(SB2),\,this work}\\
NGC 346-29&\text{MPG} 370&00:59:01.86&-72:10:43.1&15.03
&\text{O9.5\,V\,(1-7)}&\text{Binary\,(SB1)\,(7)}\\
NGC 346-31&\text{MPG} 655&00:59:15.48&-72:11:11.4&14.82
&\text{O6\,V\,(1); O5\,V+\,OB\,(5); }\\
& & & & & \text{OC5-6\,Vz\,(7)}\\
NGC 346-32&\text{MPG} 549&00:59:09.80&-72:10:58.8&15.26
&\text{O8\,V\,(1); O9.5\,V\,(7)}\\
NGC 346-35\tablefootmark{(1,2)}&\text{MPG} 508&00:59:08.11&-72:10:45.1&15.51& \hspace{0.8cm}\textemdash{} & \hspace{0.8cm}\textemdash{} &\\
NGC 346-37&\text{MPG} 593&00:59:11.61&-72:09:57.3&14.95
&\text{O5.5\,V\,(1); O8\,V\,(3)}&\text{High Mass X-ray Binary\,(6)}\\
NGC 346-38&\text{MPG} 330&00:58:58.74&-72:10:51.1&15.20
&\text{O7.5\,V\,(1); O9\,V\,(7)}\\
NGC 346-41&\text{MPG} 455&00:59:05.41&-72:10:42.1&15.25
&\text{O9.5\,V\,(1-5); B0\,V\,(7)}\\
NGC 346-44&\text{MPG} 500&00:59:07.59&-72:10:48.1&15.33
&\text{O6\,V\,(1); O8\,Vn\,(7)}\\
NGC 346-48&\text{MPG} 445&00:59:04.76&-72:11:02.7&15.33
&\text{O8\,V\,(1)}&\text{Binary\,(SB1)\,(7)}\\
NGC 346-54&\text{MPG} 356&00:59:00.92&-72:11:09.0&15.50
&\text{O6.5\,V\,(1); O6.5\,Vz\,(7)}\\
NGC 346-56&\text{MPG} 561&00:59:10.43&-72:10:46.8&15.71
&\text{O8\,V\,(1)}&\\
NGC 346-57&\text{MPG} 523&00:59:08.65&-72:10:13.9&15.50
&\text{O7\,Vz\,(3)}&\\
NGC 346-58&\text{MPG} 375&00:59:02.02&-72:10:36.0&15.47
&\text{O9.5\,V\,(8)}&\\
NGC 346-61&\text{MPG} 557&00:59:10.25&-72:10:42.4&15.89
&\text{O9.5\,V\,(1-7)}&\\
NGC 346-75\tablefootmark{(1)}&\text{MPG} 519&00:59:08.50&-72:11:12.4&16.05
&\text{O9 V\,(1-3)}&Binary\,(SB2),\,this work\\
NGC 346-76\tablefootmark{(1,2)} &\text{MPG} 427&00:59:04.20&-72:10:25.2&15.97
& \hspace{0.8cm}\textemdash{} & \hspace{0.8cm}\textemdash{} &\\
NGC 346-77&\text{MPG} 529&00:59:08.92&-72:11:10.2&16.11&\text{O7.5\,V((f))\,(1)}&\\
NGC 346-83&\text{MPG} 429&00:59:04.23&-72:10:26.9&16.01
&\text{O9.5\,V\,(8)}&\\
NGC 346-85\tablefootmark{(2)}&\text{MPG} 371&00:59:01.87&-72:10:41.4&15.60
& \hspace{0.8cm}\textemdash{} & \hspace{0.8cm}\textemdash{} &\\
NGC 346-96\tablefootmark{(1,2)} &\text{MPG} 366&00:59:01.64&-72:10:43.6&15.77
& \hspace{0.8cm}\textemdash{} & \hspace{0.8cm}\textemdash{} &\\
NGC 346-1034&\text{Star} 1034&00:59:12.81&-72:10:52.3&15.02
&\text{O9.5-B0\,V\,(7)}&\\
NGC 346-1051\tablefootmark{(1)}&\text{Star} 1051&01:00:13.61&-72:12:44.6&15.22
&\text{O9\,V\,(7)}&Binary\,(SB2),\,this work\\
NGC 346-1067&\text{Star} 1067&00:58:22.90&-72:17:51.4&15.53
&\text{O9\,V\,(7)}&\\
NGC 346-1094&\text{Star} 1094&00:59:15.87&-72:11:10.7&15.76
&\text{O8\,V\,(7)}&\\
NGC 346-1114&\text{Star} 1114&01:00:18.26&-72:07:51.8&15.98
&\text{O9\,V\,(7)}&\\
NGC 346-1144&\text{Star} 1144&00:59:11.66&-72:14:24.7&16.23
&\text{O9.5\,V\,(7)}&\\
\noalign{\smallskip}
\hline
\end{tabular}
\tablefoot{\tablefootmark{(1)}{Stars not considered in our analysis.}
\tablefootmark{(2)} Stars that had no previous spectral classification (see Appendix.,~\ref{ch:clasificacion}).}
Cross-references to identifiers are listed in the second column: MPG \citep{massey1989stellar}, Star \# \citep{dufton2019census}. 
\tablebib{(1)\cite{massey1989stellar}; (2)\cite{walborn2000ultraviolet}; (3)\cite{evans2006vlt}; (4)\cite{mokiem2006vlt}; (5)\cite{heydari2010very}; (6)\cite{antoniou2019deep}; (7)\cite{dufton2019census}; (8)\cite{rickard2022stellar}.} It is worth noting that previous works before (8) do not consider the quantitative classification criteria for the O\,Vz spectral subclass.

\end{table}

\vspace{1cm}
\begin{longtable}{cclccclcl}
\caption{\label{Table:Table_5} Estimated equivalent widths (with their errors) of the spectral lines of He I and He II, along with the assigned parameter z to identify the spectral feature Vz in the analyzed stars in 30~Doradus.}\\
\hline\hline
\text{Star} & \multicolumn{2}{c}{\small{He I $\lambda$4471}} & \multicolumn{2}{c}{\small{He II $\lambda$4542}} & \multicolumn{2}{c}{\small{He II $\lambda$4686}} & \text{z} & \text{Spectral Type} \\
            & \small{EW} & \small{$\sigma$} & \small{EW} & \small{$\sigma$} & \small{EW} & \small{$\sigma$} &           &                   \\
\hline
\endfirsthead
\caption{continued.}\\
\hline\hline
\text{Star} & \multicolumn{2}{c}{\small{He I $\lambda$4471}} & \multicolumn{2}{c}{\small{He II $\lambda$4542}} & \multicolumn{2}{c}{\small{He II $\lambda$4686}} & \text{z} & \text{Spectral Type} \\
            & \small{EW} & \small{$\sigma$} & \small{EW} & \small{$\sigma$} & \small{EW} & \small{$\sigma$} &           &                   \\
\hline
\endhead
\hline
\endfoot
14&0.61 &0.01&0.54 &0.01&0.75 &0.01&1.23&O8.5\,Vz\\
65&0.64 &0.02&0.59 &0.02&0.83 &0.02&1.30&O8\,V(n)z\\
67&0.62 &0.02&0.31 &0.02&0.66 &0.02&1.07&O9.5\,V\\
74&0.89 &0.02&0.41 &0.02&0.66 &0.02&0.74&O9\,Vn\\
89&0.54 &0.01&0.83 &0.01&1.03 &0.02&1.24&O6.5\,V((f))z\, Nstr\\
96&0.48 &0.01&0.77 &0.01&0.77 &0.01&1.00&O6\,V((n))((fc))\\
110&0.31 &0.01&0.52 &0.01&0.72 &0.01&1.39&O6\,V((n))z\\
117&0.41 &0.01&0.78 &0.02&0.99 &0.02&1.27&O6\,Vz\\
123&0.46 &0.01&0.85 &0.01&1.08 &0.01&1.27&O6.5\,Vz\\
130&0.73 &0.02&0.38 &0.01&0.66 &0.02&0.90&O8.5\,V((n))\\
132&0.68 &0.01&0.45 &0.01&0.72 &0.01&1.05&O9.5\,V\\
138&3.01 &0.02&0.40 &0.01&0.67 &0.01&0.22&O9\,Vn\\
149&0.82 &0.02&0.40 &0.02&0.69 &0.02&0.84&O9.5\,V\\
154&0.77 &0.01&0.59 &0.01&0.72 &0.01&0.94&O8.5\,V\\
168&0.73 &0.01&0.69 &0.01&0.92 &0.01&1.26&O8.5\,Vz\\
169&0.11 &0.01&0.63 &0.01&0.33 &0.01&0.53&O2.5\,V(n)((f*))\\
216&0.13 &0.01&0.73 &0.01&0.72 &0.01&0.99&O4\,V((fc))\\
249&0.73 &0.01&0.63 &0.01&0.81 &0.01&1.06&O8\,Vnz\\
250&0.78 &0.01&0.39 &0.01&0.67 &0.01&0.85&O9.2\,V((n))\\
252&0.65 &0.01&0.65 &0.01&0.90 &0.01&1.38&O8.5\,Vz\\
266&0.64 &0.01&0.69 &0.01&0.79 &0.01&1.14&O8\,V((f))z\\
280&0.81&0.01&0.45 &0.01&0.66 &0.01&0.82&O9\,V((n))\\
285&0.75 &0.01&0.61 &0.01&0.78 &0.01&1.03&O7.5\,Vnnn\\
355&0.19 &0.01&0.76 &0.01&0.78 &0.01&1.03&O4\,V((n))((fc))\\
356&0.48 &0.01&0.71 &0.01&0.87 &0.01&1.24&O6\,V(n)z\\
361&0.49 &0.01&0.43 &0.01&0.55 &0.01&1.14&O8.5\,Vz\\
369&0.49 &0.01&0.13 &0.01&0.42 &0.01&0.86&O9.7\,V\\
380&0.25 &0.01&0.64 &0.01&0.90 &0.01&1.42&O6-7\,Vz\\
382&1.45& 0.01&0.74 &0.01&0.89 &0.01&0.62&O4-5\,V((fc))\\
385&0.19 &0.01&0.71 &0.01&0.56 &0.01&0.78&O4-5\,V((n))((fc))\\
392&0.29 &0.01&0.62 &0.01&0.77 &0.01&1.24&O6-7\,V((f))z\\
398&0.39 &0.01&0.64 &0.01&0.74 &0.01&1.16&O5.5\,V((n))((f))z\\
418&0.24 &0.01&0.78 &0.02&0.93 &0.01&1.19&O5\,V((n))((fc))z\\
419&0.75 &0.01&0.39 &0.01&0.57 &0.01&0.75&O9\,V(n)\\
468&0.18 &0.01&0.64 &0.01&0.54 &0.01&0.84&O2\,V((f*))+OB\\
470&0.18 &0.01&0.70 &0.01&0.85 &0.01&1.22&O6\,V((f))z\\
472&0.28 &0.01&0.80 &0.01&0.94 &0.01&1.17&O6\,Vz\\
483&0.66 &0.01&0.23 &0.01&0.39 &0.01&0.59&O9\,V\\
484&0.47 &0.01&0.47 &0.01&0.40 &0.01&0.86&O6-7\,V((n))\\
488&0.32 &0.01&0.72 &0.01&0.78 &0.01&1.07&O6\,V((f))\\
491&0.34 &0.01&0.72 &0.01&0.70 &0.01&0.97&O6\,V((fc))\\
493&0.57 &0.01&0.45 &0.01&0.57 &0.01&1.01&O9\,V\\
494&0.67 &0.02&0.66 &0.02&0.85 &0.01&1.27&O8\,V(n)z\\
506&0.22 &0.01&0.78 &0.01&0.62 &0.01&0.80&ON2\,V((n))((f*))\\
498&0.98 &0.01&0.21 &0.02&0.38 &0.01&0.39&O9.5\,V\\
511&0.30 &0.01&0.90 &0.01&0.96 &0.01&1.07&O5\,V((n))((fc))\\
521&0.74 &0.01&0.43 &0.01&0.61 &0.01&0.82&O9\,V(n)\\
536&0.33 &0.02&0.67 &0.01&0.83 &0.02&1.24&O6\,Vz\\
537&0.17 &0.01&0.64 &0.01&0.44 &0.01&0.70&O5\,V((fc))\\
549&0.46 &0.01&0.67 &0.02&0.92 &0.02&1.38&O6.5\,Vz\\
550&2.11 &0.01&0.78 &0.01&0.82 &0.01&0.39&O5\,V((fc))\\
554&0.65 &0.02&0.24 &0.02&0.41 &0.01&0.63&O9.7\,V\\
560&0.63 &0.01&0.17 &0.01&0.49 &0.01&0.78&O9.5\,V\\
560&0.63 &0.01&0.17 &0.01&0.49 &0.01&0.78&O9.5\,V\\
577&0.92 &0.01&0.69 &0.02&0.81 &0.01&0.88&O5\,V((fc))\\
581&1.32 &0.01&0.65 &0.01&0.71 &0.01&0.54&O4-5\,V((fc))\\
582&0.64 &0.02&0.24 &0.01&0.55 &0.02&0.86&O9.5\,V((n))\\
586&0.05 &0.01&0.81 &0.01&0.92 &0.01&1.13&O4\,V((n))((fc))z\\
592&1.03 &0.02&0.24 &0.02&0.40 &0.02&0.39&O9.5\,Vn\\
597&0.76 &0.01&0.52 &0.01&0.75 &0.01&0.99&O8-9\,V(n)\\
601&0.32 &0.01&0.76 &0.01&0.82 &0.01&1.08&O5-6\,V((n))\\
611&0.66 &0.01&0.72 &0.01&0.87 &0.01&1.19&O8\,V(n)z\\
621&1.58 &0.01&0.51 &0.01&0.62 &0.01&0.39&O2\,V((f*))\\
627&0.88 &0.01&0.24 &0.01&0.54 &0.01&0.61&O9.7\,V\\
638&0.69 &0.01&0.66 &0.01&0.91 &0.01&1.32&O8.5\,Vz\\
639&0.93 &0.01&0.25 &0.01&0.51 &0.01&0.55&O9.7\,V\\
649&0.73 &0.01&0.35 &0.01&0.61 &0.01&0.84&O9.5\,V\\
660&0.95 &0.02&0.25 &0.02&0.44 &0.01&0.46&O9.5\,Vnn\\
677&0.52 &0.02&0.27 &0.02&0.53 &0.02&1.02&O9.5\,V\\
679&0.56 &0.02&0.19 &0.01&0.36 &0.01&0.65&O9.5\,V\\
704&0.76 &0.02&0.33 &0.01&0.62 &0.02&0.81&O9.2\,V(n)\\
706&0.53 &0.01&0.74 &0.01&0.81 &0.01&1.09&O6-7\,Vnn\\
722&0.70 &0.01&0.66 &0.01&0.89 &0.01&1.28&O7\,Vnnz\\
724&0.40 &0.02&0.81 &0.04&0.83 &0.04&1.02&O7\,Vnn\\
746&0.45 &0.01&0.73 &0.01&0.63 &0.01&0.87&O6\,Vnn\\
751&0.66 &0.02&0.62 &0.02&0.89 &0.02&1.35&O7-8\,Vnnz\\
755&0.69 &0.01&0.72 &0.01&0.76 &0.01&1.06&O3\,Vn((f*))\\
761&0.66 &0.01&1.11&0.01&1.14 &0.01&1.02&O6.5\,V((n))((f))\,Nstr\\
768&0.74 &0.02&0.52 &0.02&0.64 &0.02&0.86&O8\,Vn\\
770&0.63 &0.02&0.75 &0.02&0.86 &0.02&1.15&O7\,Vnnz\\
775&0.89 &0.02&0.46 &0.02&0.70 &0.02&0.79&O9.2\,V\\
778&0.81 &0.03&0.21 &0.02&0.53 &0.03&0.66&O9.5\,V\\
797&0.50 &0.01&0.76 &0.01&0.73 &0.01&0.96&O3.5\,V((n))((fc))\\
849&0.59 &0.01&0.82 &0.01&1.03 &0.01&1.26&O7\,Vz\\
892&0.86 &0.01&0.54 &0.01&0.80 &0.01&0.93&O9\,V\\
\end{longtable}

\begin{table*}[ht!]
\caption{Stellar and wind parameters obtained from the quantitative analysis of our sample.}
\label{Table:Table_6}
\centering
\begin{tabular}{lllcrrccr}
\hline\hline
\noalign{\smallskip}

Star & SpT & LC & \makecell{$v \, \mathrm{sin} \, i$ \\ $[\rm{km\,s^{-1}}]$} & \makecell{$T_\mathrm{eff}$ \\ $[\rm{kK}]$} & \makecell{$\log {\mathrm{g}}$ \\ $[\rm{dex}]$} & \makecell{$\log {\mathrm{Q}}$ \\ $[\rm{dex}]$} & \makecell{\rm{Y(He)} \\ $[\rm{dex}]$} & \makecell{\rm{$\xi_t$} \\ $[\rm{km\,s^{-1}}]$} \\

\noalign{\smallskip}
\hline
\noalign{\smallskip}
\text{NGC 346-10}& \text{O8} & V & 93& 40.8 $\pm$ 1.5& 3.79 $\pm$ 0.15 & < -14.73& 0.15 $\pm$ 0.05 & < 1.0\\
\text{NGC 346-13}& \text{O8} & Vz & 82& 37.5 $\pm$ 1.7 & 4.07 $\pm$ 0.33 & < -15.05 & 0.09 $\pm$ 0.03 & < 1.0\\
\text{NGC 346-15}& \text{O6.5} & V & 119 & 40.1 $\pm$ 0.9& 4.15 $\pm$ 0.14 & < -14.61 & 0.12 $\pm$ 0.02 & < 9.0\\
\text{NGC 346-17}&\text{O9.2} & V & 154 &32.9 $\pm$ 0.7  & 3.62 $\pm$ 0.09 & -13.22 $\pm$ 0.42 & 0.06 & < 1.0\\
\text{NGC 346-20}& \text{O7} & V & 50 &38.8 $\pm$ 1.0  & 3.87 $\pm$ 0.10& < -13.89 & 
0.10 $\pm$ 0.02 & < 1.0\\
\text{NGC 346-21}& \text{O7} & Vz & 257 &38.5 $\pm$ 1.0 & 3.96 $\pm$ 0.15&< -14.11 & 
0.14 $\pm$ 0.03 & < 1.0\\
\text{NGC 346-23}&\text{O7.5} & Vz & 50 &38.1 $\pm$ 1.1&  4.05 $\pm$ 0.15& < -13.96& 0.12 $\pm$ 0.03 & < 9.0\\
\text{NGC 346-29}& \text{O9.5} & V & 131 &34.6 $\pm$ 0.9  & 4.07 $\pm$ 0.15& 13.03 $\pm$ 0.17 & 0.10 $\pm$ 1.6 & < 1.0\\
\text{NGC 346-31}&\text{O5.5} & Vz & 90 &42.0 $\pm$ 1.0& 3.99 $\pm$ 0.10 & < -14.69 &
0.10 $\pm$ 0.02 & < 9.0\\
\text{NGC 346-32}&\text{O9.5} & V & 50 &33.3 $\pm$ 1.0 & 3.65 $\pm$ 0.15& < -15.02 &
0.11 $\pm$ 0.04 & < 9.0\\
\text{NGC 346-37}& \text{O8} & V & 185 &38.4 $\pm$ 1.2& 3.94 $\pm$ 0.13& < -13.08 &
9.4 $\pm$ 1.6 & < 1.0\\
\text{NGC 346-38}& \text{O9} & V & 151 &36.0 $\pm$ 1.3& 4.00 $\pm$ 0.20& < -15.01 &
0.13 $\pm$ 0.04 & < 5.0\\
\text{NGC 346-41}& \text{O9.5} & V & 109 &34.1 $\pm$ 0.5& 3.99 $\pm$ 0.09& < -13.84 &
0.13 $\pm$ 0.03 & < 1.0\\
\text{NGC 346-44}& \text{O8} & Vn & 280 &38.1 $\pm$ 1.2& 4.11 $\pm$ 0.18& < -14.03 &
0.15 $\pm$ 0.04 & < 20.0\\
\text{NGC 346-48}& \text{O8} & V & 197 &33.8 $\pm$ 0.9& 3.56 $\pm$ 0.16& < -14.03&
0.14$\pm$ 0.04 & < 9.0\\
\text{NGC 346-54}& \text{O6.5} & V & 50 &41.3 $\pm$ 1.2 & 4.18 $\pm$ 0.17& < -13.9 &
0.21 $\pm$ 0.06 & < 1.0\\
\text{NGC 346-56}& \text{O8} & V & 96 &38.2  $\pm$ 1.3& 4.14 $\pm$ 0.26& < -14.06 &
0.14 $\pm$ 0.04 & < 1.0\\
\text{NGC 346-57}& \text{O7} & Vz & 50 &39.0 $\pm$ 1.0 & 3.96 $\pm$ 0.15& < -14.68 &
0.12 $\pm$ 0.03 & < 5.0\\
\text{NGC 346-58}& \text{O9.5} & V & 224 &34.5 $\pm$ 1.1 & 4.08 $\pm$ 0.21& < -15.01 &
0.14 $\pm$ 0.04 & < 1.0\\
\text{NGC 346-61}& \text{O9.5} & V & 83 &32.9 $\pm$ 1.4 & 3.62 $\pm$ 0.23& < -14.01 &
0.10 & < 1.0 \\
\text{NGC 346-77}&\text{O7.5} & Vz & 50 &37.3 $\pm$ 1.4 & 3.77 $\pm$ 0.20& < -13.99 &
0.13 $\pm$ 0.04 & < 1.0\\
\text{NGC 346-83}& \text{O9.5} & V & 50 &33.3 $\pm$ 0.9 & 3.97 $\pm$ 0.20& < -14.01&
0.12 $\pm$ 0.03 & < 5.0\\
\text{NGC 346-85}&\text{O9.7} & V & 108 &32.2 $\pm$ 0.7 & 3.85 $\pm$ 0.13& 12.77 $\pm$ 0.23 & 0.09 $\pm$ 0.02 & < 1.0\\
\text{NGC 346-1034}& \text{O9.5} & V & 136 &35.0 $\pm$ 1.3 & 4.19 $\pm$ 0.23& 12.64 $\pm$ 0.22 & 0.06 & < 1.0\\
\text{NGC 346-1067}& \text{O8.5} & Vz & 94 &36.6 $\pm$ 1.6 & 4.05 $\pm$ 0.28& < -15.03 &
0.11 $\pm$ 0.05 & < 5.0\\
\text{NGC 346-1094}&\text{O8} & Vz & 77 &37.2 $\pm$ 1.4 & 4.05 $\pm$ 0.21& < -12.94 &
0.10 & <1.0 \\
\text{NGC 346-1114}& \text{O9} & V & 236 &35.7 $\pm$ 1.7 & 4.14 $\pm$ 0.29& < -14.06 &
0.11 $\pm$ 0.04 & < 1.0\\
\text{NGC 346-1144}&\text{O9.5} & V & 50 &35.3 $\pm$ 1.4 & 4.07 $\pm$ 0.27& < -14.73 &
0.10 $\pm$ 0.04 & < 1.0\\
\noalign{\smallskip}
\hline
\end{tabular}
\end{table*}

\onecolumn
\section{Figures} \label{ch:Figures}
\begin{figure}[ht!]
    \centering
    \includegraphics[width=\linewidth, height=0.90\textheight]{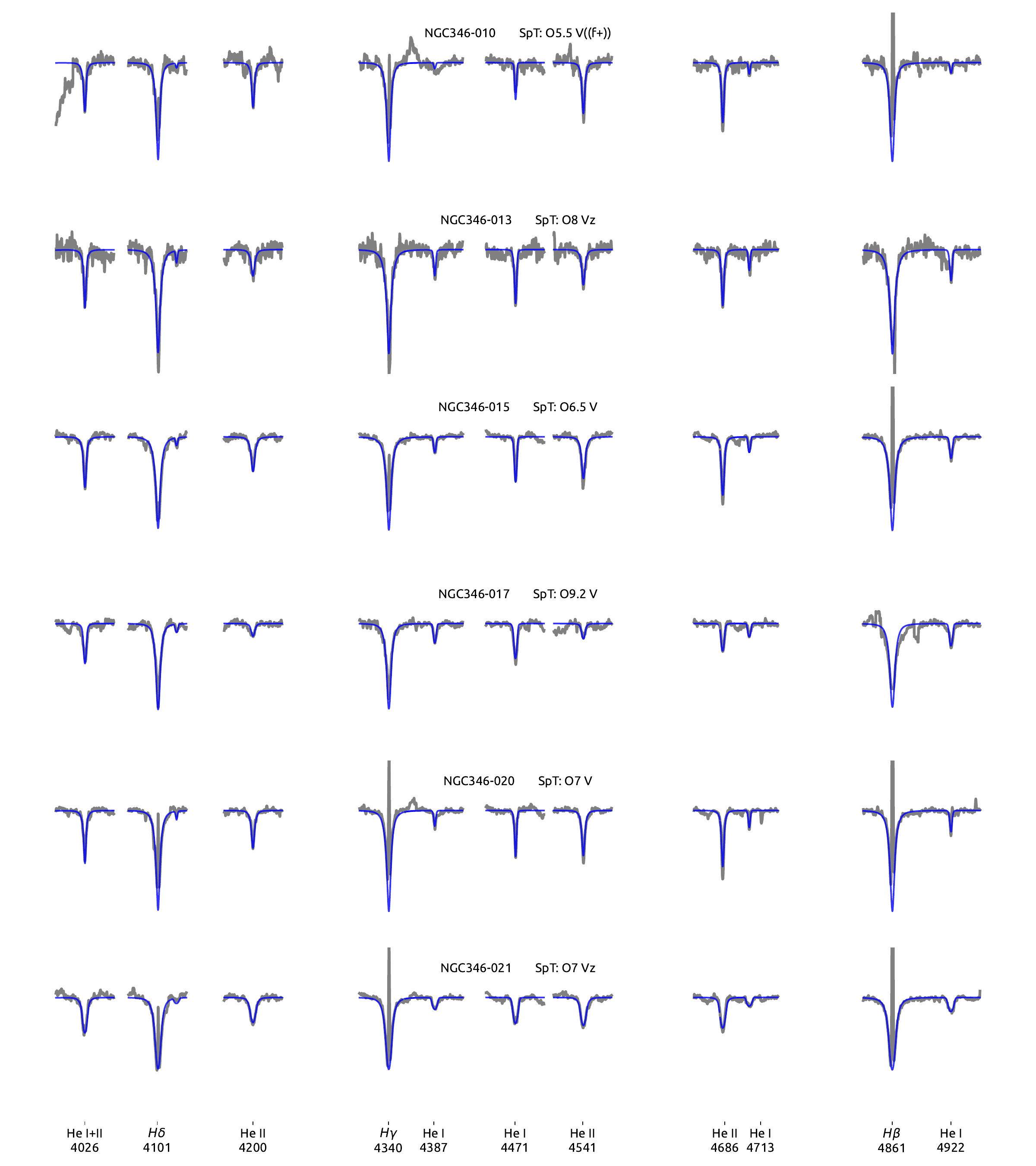}
    \caption{Observed spectra (grey) and best fitting model (blue) comparison for the main diagnostic lines. The sample is sorted following Table~\ref{Table:Table_6}.}
    \label{Figure:Figure_12}
\end{figure}

\begin{figure}[ht!]
    \ContinuedFloat
    \centering
    \includegraphics[width=\linewidth, height=0.95\textheight]{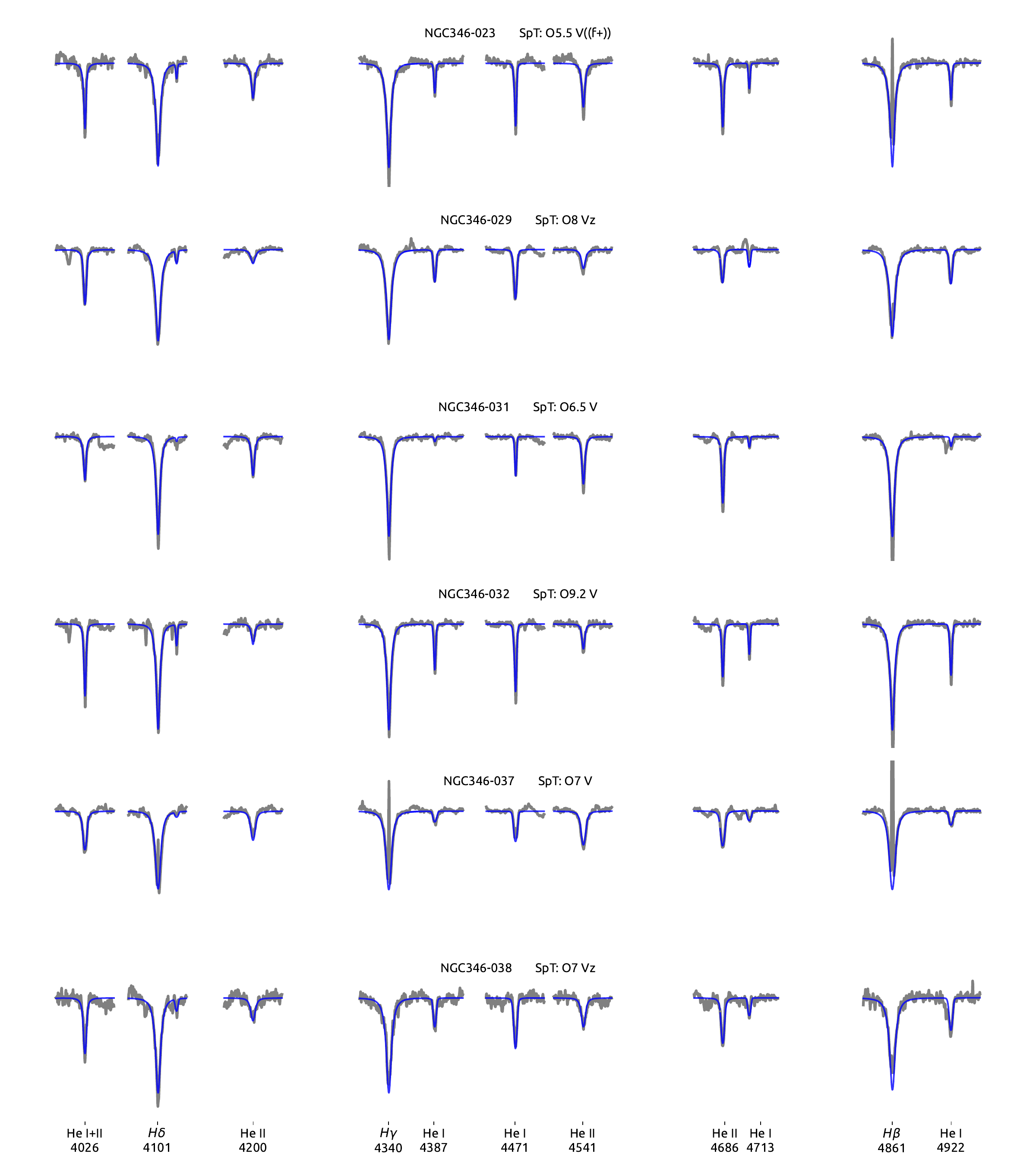}
    \caption*{Fig.~\ref{Figure:Figure_12} continued.}
\end{figure}

\begin{figure}[ht!]
    \ContinuedFloat
    \centering
    \includegraphics[width=\linewidth, height=0.95\textheight]{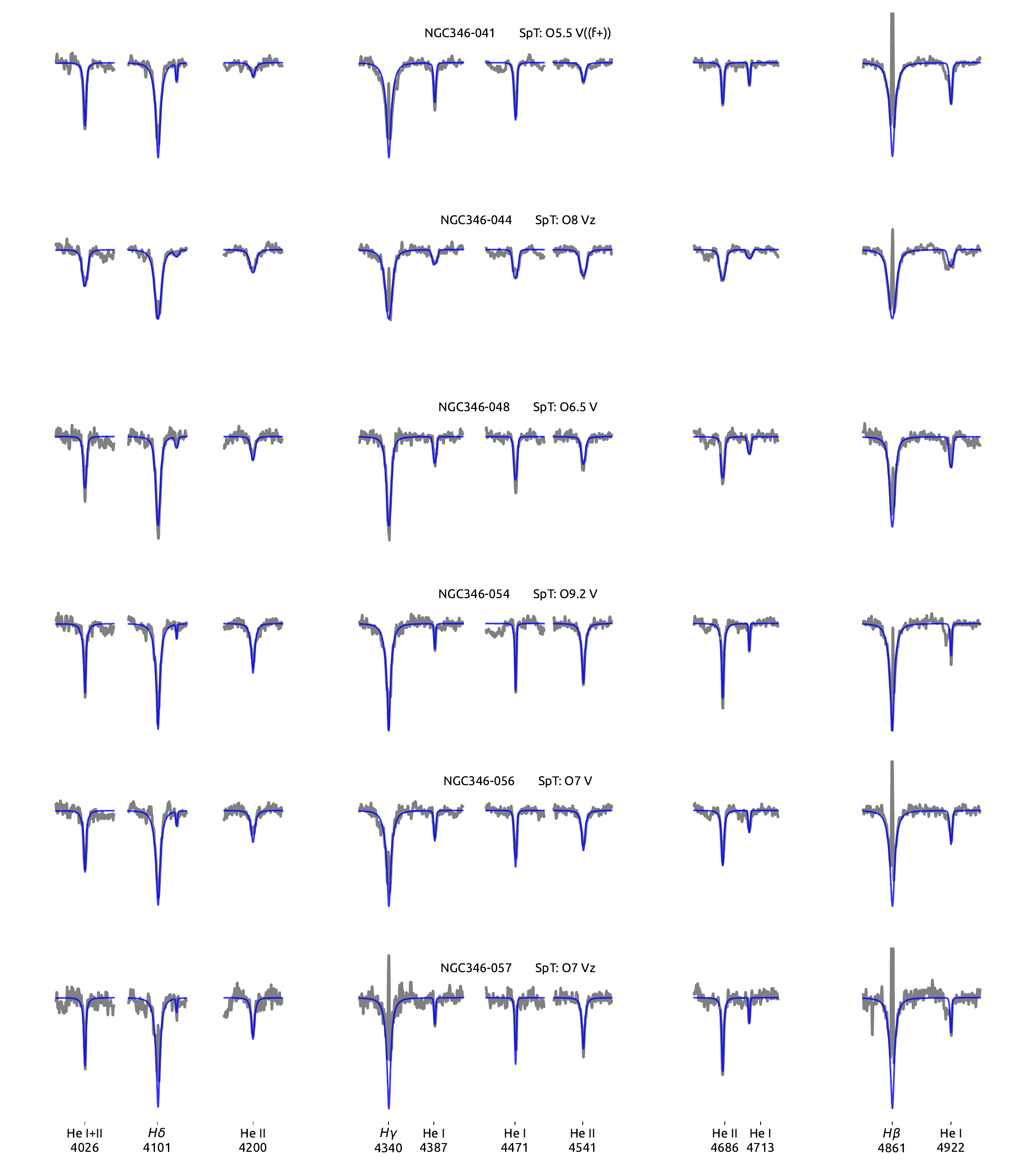}
    \caption*{Fig.~\ref{Figure:Figure_12} continued.}
\end{figure}

\begin{figure}[ht!]
    \ContinuedFloat
    \centering
    \includegraphics[width=\linewidth, height=0.95\textheight]{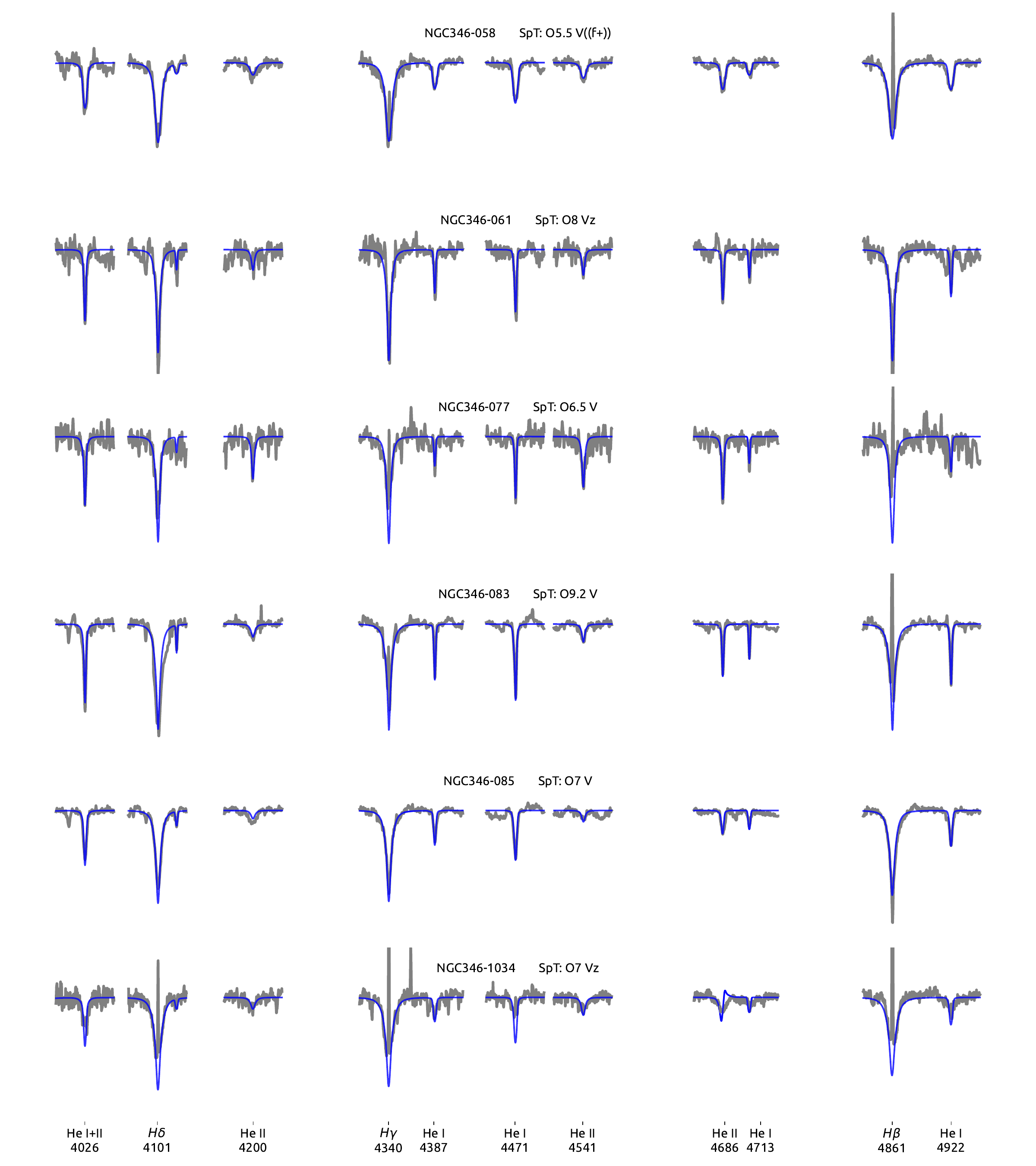}
    \caption*{Fig.~\ref{Figure:Figure_12} continued.}
\end{figure}

\begin{figure}[ht!]
    \ContinuedFloat
    \centering
    \includegraphics[width=\linewidth, height=0.60\textheight]{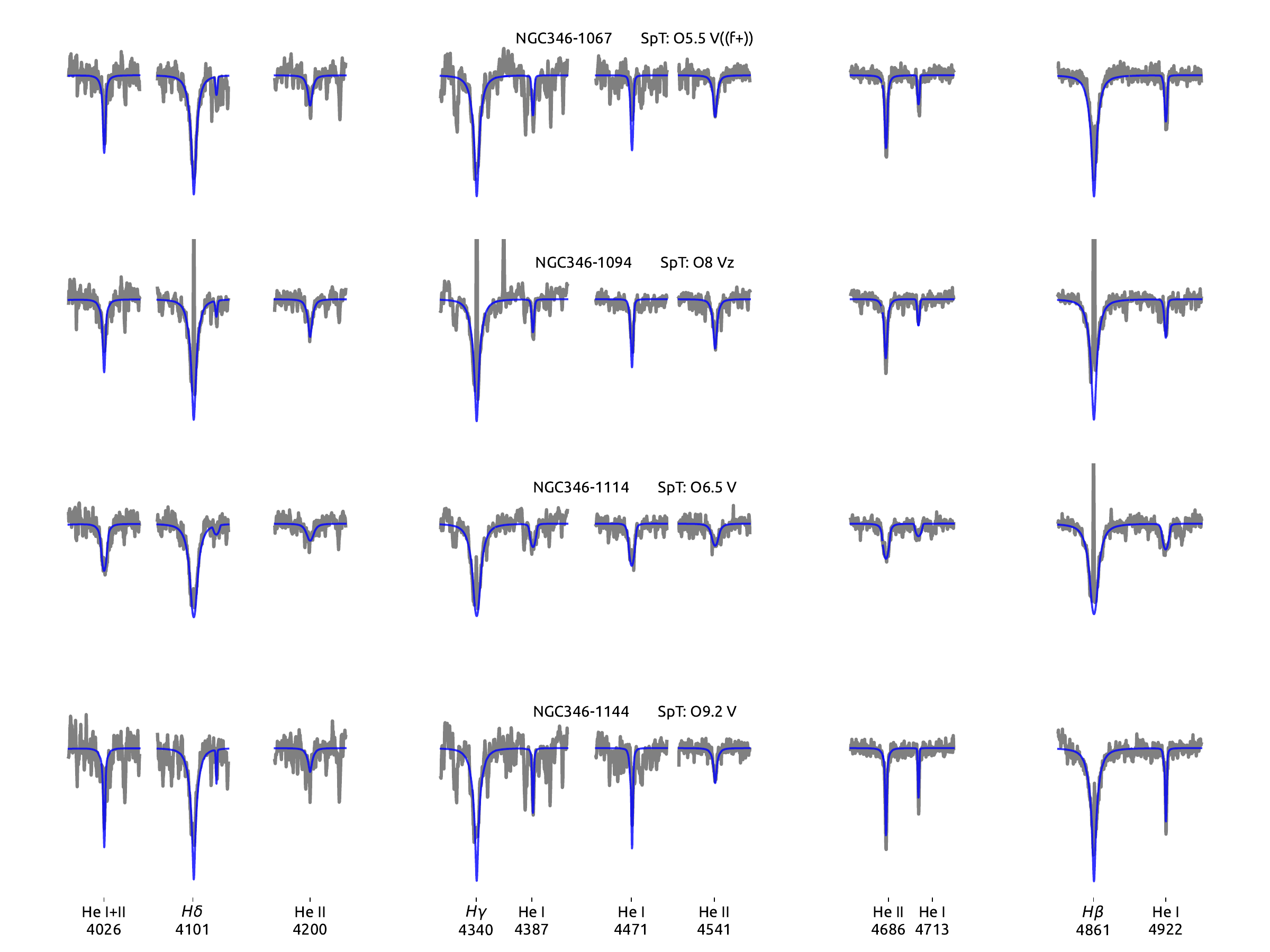}
    \caption*{Fig.~\ref{Figure:Figure_12} continued.}
\end{figure}

\end{appendix}
\end{document}